\documentclass[10pt,amsmath,amssymb,prl,superscriptaddress, twocolumn]{revtex4}
\usepackage{bm}
\usepackage[dvipdfmx]{graphicx, color}
\newcounter{one}
\newcommand{\bra}[1]{\langle #1 |}
\newcommand{\ket}[1]{| #1 \rangle}
\newcommand{\ketbra}[1]{| #1 \rangle\langle #1 |}

\newcommand{\Tr}[0]{ \textrm{Tr}}

\newcommand{\ca}[1]{{\cal #1}}
\newtheorem{theorem}{Theorem}
\newtheorem{lemma}{Lemma}
\newtheorem{definition}{Definition}
\newtheorem{corollary}{Corollary}

\def\QED{\mbox{\rule[0pt]{1.5ex}{1.5ex}}}

\def\endproof{\hspace*{\fill}~\QED\par\endtrivlist\unskip}

\newenvironment{proofof}[1]{\vspace*{5mm} \par \noindent
         {\bf Proof of #1:\hspace{2mm}}}{\endproof
}

\def\Label{\label}

\newcommand{\affA}{
Center for Emergent Matter Science (CEMS), RIKEN, Wako, Saitama 351-0198 Japan
}
\newcommand{\affB}{Graduate School of Information Systems, The University of Electro-Communications, Chofu, Japan}

\begin{document}
\title{Large Deviation implies First and Second Laws of Thermodynamics}
\author{Hiroyasu Tajima}
\affiliation{\affA}
\author{Eyuri Wakakuwa}
\author{Tomohiro Ogawa}
\affiliation{\affB}

\begin{abstract}
To reconstruct thermodynamics based on the microscopic laws is one of the most important unfulfilled goals of statistical physics.
Here, we show that the first law and the second law for adiabatic processes are derived from an assumption that ``probability distributions of energy in Gibbs states satisfy large deviation'', which is widely accepted as a property of thermodynamic equilibrium states.
We define an adiabatic transformation as a randomized energy-preserving unitary transformations on the many-body systems and the work storage.
As the second law, we show that an adiabatic transformation from a set of Gibbs states to another set of Gibbs states is possible if and only if the regularized von Neumann entropy becomes large. 
As the first law, we show that the energy loss of the thermodynamic systems during the adiabatic transformation is stored in the work storage as ``work,'' in the following meaning;
(i) the energy of the work storage takes certain values macroscopically, in the initial state and the final state.
(ii) the entropy of the work storage in the final state is macroscopically equal to the entropy of the initial state. 
As corollaries, our results give the principle of maximam work and the first law for the isothermal processes.
\end{abstract}
 
\maketitle

Statistical mechanics was born to reconstruct thermodynamics from the microscopic mechanical laws \cite{Boltzmann}.
It has achieved great results and succeeded in elucidating various nonequilibrium phenomena beyond thermodynamic region \cite{Landau}.
However, the original goal of statistical mechanics, "Reconstructing thermodynamics from microscopic mechanical laws" has not been fully accomplished \cite{Lieb,Shimizu}.
In thermodynamics, whether macroscopic state transformation is possible or not is completely determined by the values of the thermodynamic functions of the initial and final states of the transformation \cite{Lieb,Shimizu}.
In the case of adiabatic processes, a combination of the equilibrium states $(T_{1},V_{1},N_{1};...;T_{m},V_{m},N_{m})$ can be adiabatically transformed to another combination $(T'_{1},V'_{1},N_{1};...;T'_{m},V'_{m},N_{m})$ if and only if the following inequality holds:
\begin{align}
\Delta S_{S}\ge0,
\end{align}
where $\Delta S_{S}$ is the difference of the thermodynamic entropy of the combinations $(T_{1},V_{1},N_{1};...;T_{m},V_{m},N_{m})$ and $(T'_{1},V'_{1},N_{1};...;T'_{m},V'_{m},N_{m})$, i.e, $\Delta S_{S}:=\sum_{k}S(T'_{k},V'_{k},N_{k})-S(T_{k},V_{k},N_{k})$.
The energy lost from the thermodynamic systems during the adiabatic transformation is stored in the external work storage in a special form called ``work:''
\begin{align}
W=-\Delta U_{S}
\end{align}
These are the first law and the second law of thermodynamics. They are the two principles which form the basis of thermodynamics.

In order to derive these two laws from microscopic mechanical laws, much effort has been done in the field of statistical mechanics \cite{Landau,Jarzynski,tasaki,Kurchan,Car1,Xiao,tasaki15,Lenard,Croocks}.
However, the reconstruction of these laws is still not fully achieved.
Regarding the second law, although these results have been successfully demonstrated to show the direct part ``the inequality holds if the conversion is possible,"
the converse part ``the conversion is possible if the inequality holds" has not been shown yet.
Regarding the first law, although numerous studies \cite{Sekimoto, Max} try to distinguish a special energy form ``work'' from another form ``heat'' statistically mechanically, a complete conclusion has not been obtained yet.

Recently, the above problems in reconstruction of the two laws are rapidly being solved by the approach from quantum information theory \cite{Horodecki,oneshot1,oneshot3,Egloff,Brandao,Car2,Popescu2014,Popescu2015,oneshot2,Weilenmann2015,review,catalyst,Malabarba,optimal,m-based,T-W}.
These results give many necessary and sufficient conditions for the possibility of the transformation caused by the interaction with the heat baths.
The conditions can be interpreted as extensions of the second law into nanoscale, and are identical with the second law of thermodynamics when the size of the system is infinitely large. 
In spite of the splendid success, however, the following open problems remain:
\begin{description}
\item1. The results are based on strong assumptions about thermodynamic systems and heat baths, e.g., the i.i.d. feature and/or the number of degeneracy. 
The assumptions are not necessarily satisfied by actual thermodynamic systems.
Ref. \cite{T-W} shows that the second law in $(U,V,N)$ expression is derived from a weak assumption that the regularized Boltzmann entropy exists, but the second law in $(T,V,N)$ expression is still remain.
\item2. The analysis is mainly limited to cases where the baths are sufficiently large compared to the system, so that the temperatures of baths are almost unchanged during the thermal processes.
In the adiabatic transformation without such large heat baths, the temperatures of all systems greatly change. 
Understanding how such temperature changes occur is still limited.
\item3. There is no consensus about the statistical mechanical definitions of ``work'' and ``heat.'' 
Some of the above studies \cite{Horodecki, oneshot1, oneshot3, Egloff, Brandao, Car2, oneshot2, Weilenmann2015} also approach to this problem, by employing the notion of the single-shot work extraction, in which the work extraction is defined as a deterministic translation from the ground pure state to the excited pure state in a two-level system.
It defines a work-like energy transfer in the quantum scale well.
However, it is not clear whether these formulations completely capture the notion of ``work'' in thermodynamics.
\end{description}

In this paper, we tackle these problems. 
We show that the first law and the second law are derived from an assumption that ``probability distributions of energy in Gibbs states satisfy large deviation'', which is widely accepted as a property of thermodynamic equilibrium states.
We treat thermodynamic systems and an external work storage as quantum systems, and define an adiabatic transformation as a probabilistic mixture of the energy-preserving unitary transformations on the thermodynamic systems and the work storage.
As the second law, we show that an adiabatic transformation from a combination of Gibbs states corresponding to $(T_{1},V_{1},N_{1};...;T_{m},V_{m},N_{m})$ to another combination of Gibbs states corresponding to $(T'_{1},V'_{1},N_{1};...;T'_{m},V'_{m},N_{m})$ is possible if and only if the regularized von Neumann entropy becomes large:
\begin{align}
\Delta \tilde{S}_{S}\ge0,
\end{align}
where $\Delta \tilde{S}_{S}$ is the difference of the regularized von Neumann entropy defined for the macroscopic parameters $(T_{1},V_{1},N_{1};...;T_{m},V_{m},N_{m})$ and $(T'_{1},V'_{1},N_{1};...;T'_{m},V'_{m},N_{m})$; $\Delta \tilde{S}_{S}:=\sum_{k}\tilde{S}(T'_{k},V'_{k},N_{k})-S(T_{k},V_{k},N_{k})$.
As the first law, we show that the energy lost from the thermodynamic systems during the adiabatic transformation is stored in the work storage as ``work,'' in the following meaning;
(i) the energy of the work storage takes certain values macroscopically, in the initial state and the final state.
(ii) the entropy of the work storage in the final state is macroscopically equal to the entropy of the initial state.

As corollaries, our results give other forms of the first law and the second law.
As an example, we give the maximam work principle $\tilde{W}\le-\Delta \tilde{F}_{S}$ and the first law $\tilde{W}=-\Delta \tilde{U}_{S}+\tilde{Q}$ in the isothermal processes.

\section{Formulation}

In this section, we introduce the formulation treated in the present paper.
The system we deal with is divided into two parts, i.e., the internal system $I$ and the external work storage $E_{W}$.
The internal system is the composite system of the thermodynamic systems and the control system.
We firstly explain the internal system, secondly explain the external work storage, and finally explain the dynamics on $IE_{W}$.

\textit{Internal system;}
We consider the internal system $I$ as the composite system of the thermodynamic systems $S_{1}$,...,$S_{m}$ and the control system $C$.
We refer to the Hilbert spaces of $S_{1}$,...,$S_{m}$ and $C$ as $\ca{H}^{(n)}_{S_{1}}$,...,$\ca{H}^{(n)}_{S_{m}}$ and $\ca{H}_{C}$, respectively.
Here, $n$ is the scaling parameter, and the macroscopic limit is defined as the limit of $n\rightarrow\infty$.
We consider each $S_{k}$ as a composite system of $n$ subsystems, whose dimension is $d_{k}$.
For example, when $S_{k}$ is a compsite system of $n$ $d_{k}$-level spins,  $d_{k}$ is finite, and the dimension of $\ca{H}^{(n)}_{S_{k}}$ is $d^{n}_{k}$.
When $S_{k}$ is a compsite system of $n$ harmonic oscilators, $d_{k}$ is infinite, and the dimension of $\ca{H}^{(n)}_{S_{k}}$ is also infinite.
For the simplicity, we refer to the composite system of $S_{1}$...$S_{m}$ as $S$.

The Hamiltonian of the internal system $I$ is described as follows:
\begin{align}
H^{(n)}_{I}:=\sum_{\vec{\lambda}=(\lambda_{1},...,\lambda_{m})}(H^{(n)}_{1|\lambda_{1}}+...+H^{(n)}_{m|\lambda_{m}})\otimes\ket{\vec{\lambda}}\bra{\vec{\lambda}}_{C},
\end{align}
where $\vec{\lambda}:=(\lambda_{1},...,\lambda_{m})$ represents the set of the control parameters, e.g., the strength of the magnetic field or the position of the piston.
These parameters $\vec{\lambda}$ are registered in a finite-dimensional system $C$ as the state $\ket{\vec{\lambda}}$ on $\ca{H}_{C}$.
The operator $H^{(n)}_{l|\lambda_{l}}$ is the Hamiltonian on $\ca{H}^{(n)}_{S_{l}}$.
It is given as a function of the control parameter $\lambda_{l}$.
By using Hamiltonian $H^{(n)}_{I}$, we can effectively change the Hamiltonians of $S_ {1}$, ..., $S_ {m}$ by changing the state of $C$ \cite{catalyst}.
For the simplicity, we use the abbreviation $H^{(n)}_{S|\vec{\lambda}}:=H^{(n)}_{1|\lambda_{1}}+...+H^{(n)}_{m|\lambda_{m}}$.

As the equiribrium states of $S_{1}$,...,$S_{m}$, we employ Gibbs states.
We describe the Gibbs state of $S_{l}$ with the control parameter $\lambda_{l}$ and the inverse temperature $\beta_{l}$ as follows:
\begin{align}
\rho^{(n)}_{\beta_{l}|\lambda_{l}}:=\frac{\exp[-\beta_{l}H^{(n)}_{l|\lambda_{l}}]}{\Tr[\exp[-\beta_{l}H^{(n)}_{l|\lambda_{l}}]]}
\end{align}
We assume that the initial state $\rho^{(n)}_{I}$ of $I$ is as follows:
\begin{align}
\rho^{(n)}_{I}=\rho^{(n)}_{\vec{\beta}|\vec{\lambda}}\otimes\ket{\vec{\lambda}}\bra{\vec{\lambda}}_{C},
\end{align}
where we use the abbreviation $\rho^{(n)}_{\vec{\beta}|\vec{\lambda}}:=\rho^{(n)}_{\beta_{1}|\lambda_{1}}\otimes...\otimes\rho^{(n)}_{\beta_{m}|\lambda_{m}}$

Finally, we require that when the thermodynamic systems $S_{1}$, ..., $S_{m}$ are in Gibbs states, they satisfy the following large deviation principle of energy.
That is, for an arbitrary positive number $x$, the following approximations hold:
\begin{align}
\Tr[\rho^{(n)}_{\beta_{l}|\lambda_{l}}\Pi^{(n)}_{l,\ge n(\tilde{U}_{l}(\beta_{l},\lambda_{l})+x)}]&\approx \exp[-nI_{l}(x|\beta_{l},\lambda_{l})].\Label{LDf1}\\
\Tr[\rho^{(n)}_{\beta_{l}|\lambda_{l}}\Pi^{(n)}_{l,\le n(\tilde{U}_{l}(\beta_{l},\lambda_{l})-x)}]&\approx \exp[-nI_{l}(-x|\beta_{l},\lambda_{l})].\Label{LDf2}
\end{align}
where $\tilde{U}_{l}(\beta_{l},\lambda_{l})$ is the regularized internal energy
\begin{align}
\tilde{U}_{l}(\beta_{l},\lambda_{l}):=\lim_{n\rightarrow\infty}\frac{1}{n}\Tr[\rho^{(n)}_{\beta_{l}|\lambda_{l}}H^{(n)}_{l|\lambda_{l}}],
\end{align}
and where $\Pi^{(n)}_{l,\ge na}$ and $\Pi^{(n)}_{l,\le na}$ are the projection to the subspace of $\ca{H}^{(n)}_{S_{l}}$ which is spanned by the energy eigenvectors of $H^{(n)}_{l|\lambda_{l}}$ whose eigenvalues are larger than or equal to $na$, and are lower than or equal to $na$, respectively.
Moreover, $I_{l}(x|\beta_{l},\lambda_{l})$ is a convex function which becomes 0 only when $x=0$.

For Gibbs state satisfying the above assumption, the regularized Helmholtz free energy and the regularized entropy are also defined as follows:
\begin{align}
\tilde{F}_{l}(\beta_{l},\lambda_{l})&:=\lim_{n\rightarrow\infty}\frac{-\log \Tr[\exp[-\beta_{l}H^{(n)}_{l|\lambda_{l}}]]}{n\beta_{l}}\\
\tilde{S}_{l}(\beta_{l},\lambda_{l})&:=\lim_{n\rightarrow\infty}\frac{-\Tr[\rho^{(n)}_{\beta_{l}|\lambda_{l}}\log\rho^{(n)}_{\beta_{l}|\lambda_{l}}]}{n}
\end{align}
These two quantities satisfy the following relation:
\begin{align}
\tilde{S}_{l}(\beta_{l},\lambda_{l})=\beta_{l}(\tilde{U}_{l}(\beta_{l},\lambda_{l})-\tilde{F}_{l}(\beta_{l},\lambda_{l}))\Label{SUF}.
\end{align}
We assume that $\tilde{U}$ and $\tilde{S}$ are continuous and decreasing functions of $\beta$.
This assumption implies that the phase transition does not occur.
Even when dealing with a system with a phase transition, if we limit the temperature range to be handled to a region without phase transition, the results of this paper is perfectly applicable.

\textit{Work storage:}
Next, we formulate the work storage $E_{W}$ which stores the energy extracted from $I$.
As the preparation, we clarify the concept of ``work" treated in this paper.
Intuitively, it can be said that the work extraction satisfies at least the following two conditions:$\\$ 
1. Before and after the work extraction, the energy of the work storage must have definite values macroscopically.$\\$ 
2. The entropy of the work storage must not change macroscopically before and after the work extraction.

From the condition 1, by the work extraction, the state of $E_{W}$ is changed from a microcanonical state of energy (or a state close to the microcanonical state)  to another microcanonical state of energy (or a state close to the microcanonical state) .
Because of the condition 2, the entropy of the initial state of $E_{W}$ (=a microcanonical state) is equal to the entropy of the final state of $E_{W}$ (= another microcanonical state). 
Therefore, the density of states of $E_{W}$ should be constant with energy.
This condition means that $E_{W}$ is not a thermodynamic system, but a mechanical system.
 (Fig. \ref{mechanical}).

\begin{figure}
\begin{center}
\includegraphics[scale=0.3]{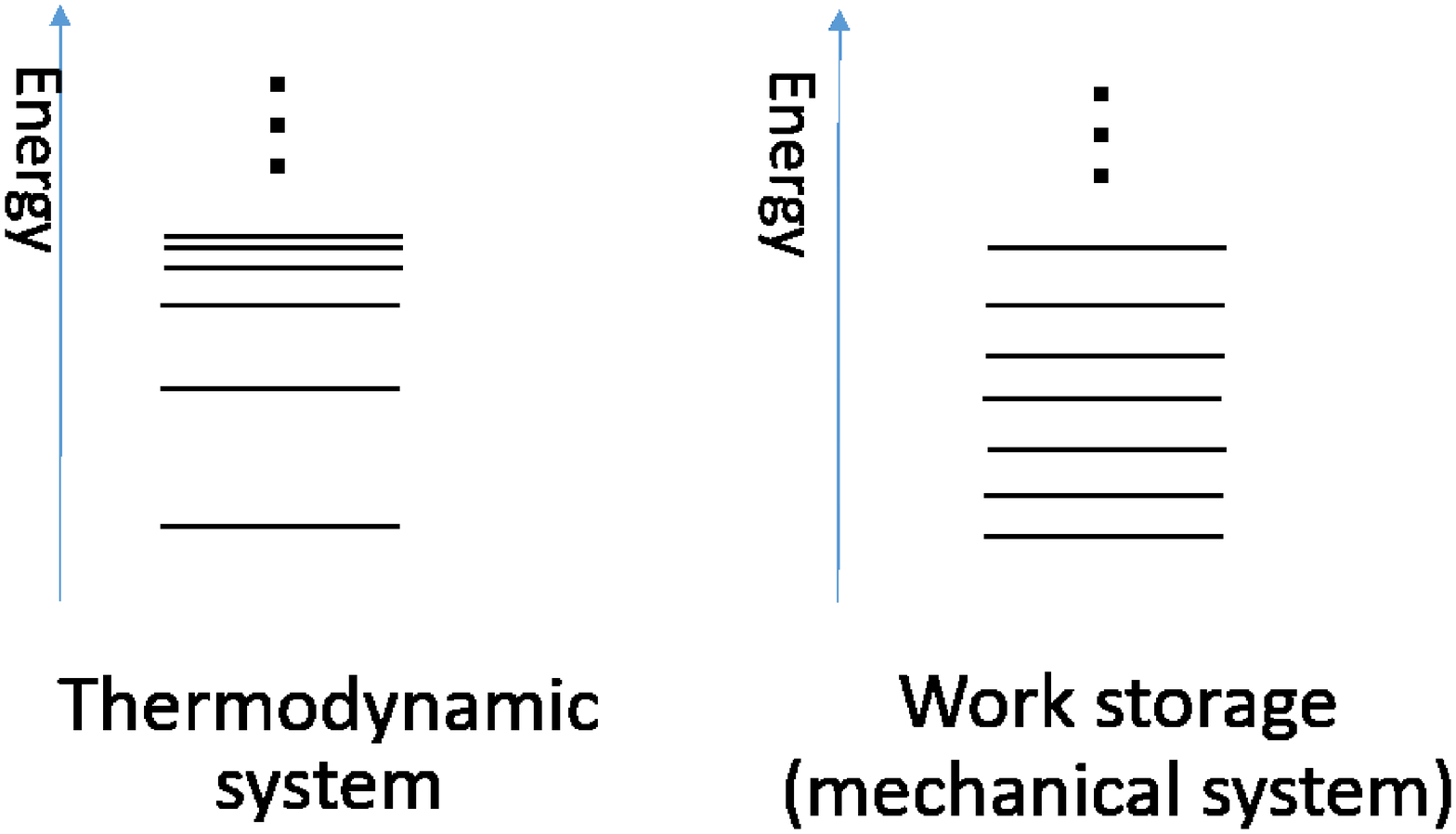}
\end{center}
\caption{Difference between thermodynamic systems and work storage.
In the thermodynamic system, the energy levels become dense as the energy rises. 
In the work storage, the density of the energy levels is independent of energy.}
\Label{mechanical}
\end{figure}

Mathematically, we require $E_{W}$ to satisfy the following three conditions:$\\$ 
(A): The Hamiltonian $\hat{H}^{(n)}_{E_{W}}:=\sum_{e\in \Lambda^{(n)}} e|e\rangle_{E_{W}}~_{E_{W}}\langle e|$ has an ground energy eigenvalue, i.e., there exists the minimum in $\Lambda^{(n)}$.
We fix the energy eigenvalue of the ground states as 0. $\\$
(B): For arbitrary energy eigenvalues $h$ and $h'$ of $H^{(n)}_{I}$ and an arbitrary $e\in \Lambda^{(n)}$ satisfying $e+h-h'>0$, there exists a $e'\in \Lambda^{(n)}$ satisfying $e'=e+h-h'$.$\\$ 
(C): The density of states of $E_{W}$ is constant with energy. Namely, for an arbitrary positive number $\delta>0$, the dimension $\Omega^{(n)}_{E_{W}}(U,\delta)$ of the following subspace of $\ca{H}^{(n)}_{E_{W}}$ is independent of $U$ whenever $U-\delta>0$ holds:
\begin{align}
\ca{H}^{(n)}_{E_{W}}(U,\delta)=\mathrm{span}\{&\ket{\psi}|H^{(n)}_{E_{W}}\ket{\psi}=\lambda\ket{\psi},\nonumber\\
& n(U-\delta)\le\lambda\le n(U+\delta)\}
\end{align}

\textit{Dynamics: randamized energy preserving unitary}
We next formulate the dynamics on $IE_{W}$ as the randomized energy-preserving unitary dynamics $\{p^{(n)}_{j},V^{(n)}_{j}\}$.
Here, $\{p^{(n)}_{j}\}$ is an arbitrary probability distribution, and $\{V^{(n)}_{j}\}$ is a set of the unitary transformations on $IE_{W}$ preserving the energy of $IE_{W}$:
\begin{align}
[V^{(n)}_{j},H^{(n)}_{I}+H^{(n)}_{E_{W}}]=0.
\end{align}

Under the above formulation, we define the work extraction by the adiabatic transformation as follows:$\\$
\textit{Adiabatic work extraction:}
Let us fix arbitrary sets of the inverse temperatures and the control parameters $(\vec{\beta},\vec{\lambda}):=(\beta_{1},\lambda_{1};...;\beta_{m},\lambda_{m})$ and $(\vec{\beta}',\vec{\lambda}'):=(\beta'_{1},\lambda'_{1};...;\beta'_{m},\lambda'_{m})$, and fix an arbitrary real number $\tilde{W}$. 
Then,  the adiabatic work extraction is defined as follows:$\\$
``The adiabatic work extraction $(\vec{\beta},\vec{\lambda}')\rightarrow_{\mathrm{ad}:\tilde{W}}(\vec{\beta}',\vec{\lambda}')$ is possible'' 
$\Leftrightarrow_{\mathrm{def}}$
``for an arbitrary positive number $\delta>0$, there exist a real positive number $\tilde{U}_{W}$ and a randomized energy-preserving unitary $\{p^{(n)}_{j},V^{(n)}_{j}\}$ satisfy the following three conditions for sufficiently large $n$:''
\begin{align}
\lim_{n\rightarrow\infty}\|\sigma^{(n)}_{I}-\rho^{(n)}_{\vec{\beta}'|\vec{\lambda}'}\otimes\ket{\vec{\lambda}'}\bra{\vec{\lambda}'}\|&=0,\Label{awe1}\\
\lim_{n\rightarrow\infty}\|\sigma^{(n)}_{E_{W}}-\pi^{(n)}_{\tilde{U}_{W}+\tilde{W},\delta}\|&=0\Label{awe3}\\
\lim_{n\rightarrow\infty}\frac{S(\sigma^{(n)}_{E_{W}})-S(\pi^{(n)}_{\tilde{U}_{W},\delta})}{n}&=0\Label{awe4}
\end{align}
where $\pi^{(n)}_{\tilde{U},\delta}$ is the maximally mixed state of $\ca{H}^{(n)}_{E_{W}}(\tilde{U},\delta)$, $\sigma^{(n)}_{IE_{W}}:=\sum_{j}p^{(n)}_{j}V^{(n)}_{j}(\rho^{(n)}_{I}\otimes\pi^{(n)}_{\tilde{U}_{W},\delta})V^{(n)\dagger}_{j}$,  $\sigma^{(n)}_{I}:=\Tr_{E_{W}}[\sigma^{(n)}_{IE_{W}}]$ and $\sigma^{(n)}_{E_{W}}:=\Tr_{I}[\sigma^{(n)}_{IE_{W}}]$.

\section{Main Result}
Under the above formulation, the following theorem holds:
\begin{theorem}\Label{Tmain}
For arbitrary sets of the parameters  $(\vec{\beta},\vec{\lambda}):=(\beta_{1},\lambda_{1};...;\beta_{m},\lambda_{m})$ and $(\vec{\beta}',\vec{\lambda}'):=(\beta'_{1},\lambda'_{1};...;\beta'_{m},\lambda'_{m})$ and an arbitrary real number $\tilde{W}$, the adiabatic work extraction $(\vec{\beta},\vec{\lambda})\rightarrow_{\mathrm{ad}:\tilde{W}}(\vec{\beta}',\vec{\lambda}')$ is possible if and only if the following two expressions hold:
\begin{align}
\Delta \tilde{S}_{S}&\ge0\\
\tilde{W}&=-\Delta \tilde{U}_{S}
\end{align}
where, $\Delta \tilde{S}_{S}:=\sum_{l}\tilde{S}_{l}(\beta'_{l},\lambda'_{l})-\sum_{l}\tilde{S}_{l}(\beta_{l},\lambda_{l})$ and $\Delta \tilde{U}_{S}:=\sum_{l}\tilde{U}_{l}(\beta'_{l},\lambda'_{l})-\sum_{l}\tilde{U}_{l}(\beta_{l},\lambda_{l})$.
\end{theorem}
Theorem \ref{Tmain} is very simple, but it is different from the usual form of the first law and the second law of thermodynamics.
Therefore, we derive the usual form of the first and the second law of thermodynamics for adiabatic processes as the corollaries of Theorem \ref{Tmain}.
Firstly, we derive the second law.
The second law of thermodynamics focus on the covertibility from a combination of equilibrium states $(\vec{\beta},\vec{\lambda})$ to another combination of equilibrium  $(\vec{\beta}',\vec{\lambda}')$ by the adiabatic transformation, without considering the amount of extracted work.
Therefore, the statement that ``the conversion $(\vec{\beta},\vec{\lambda})$ to $(\vec{\beta}',\vec{\lambda}')$ by the adiabatic transformation is possible'' is equivalent to the statement ``there exists a real number $\tilde{W}$, and the adiabatic work extraction $(\vec{\beta},\vec{\lambda})\rightarrow_{\mathrm{ad}:\tilde{W}}(\vec{\beta}',\vec{\lambda}')$.''
Hence, the second law is easily derived from Theorem \ref{Tmain}:
\begin{corollary}[Second Law for adiabatic processes]\Label{coro1}
The conversion $(\vec{\beta},\vec{\lambda})$ to $(\vec{\beta}',\vec{\lambda}')$ by the adiabatic transformation is possible if and only if $\Delta \tilde{S}_{S}\ge0$ holds.
\end{corollary}

Similarly, the first law is easily derived from Theorem \ref{Tmain}:
\begin{corollary}[First Law for adiabatic processes]\Label{coro2}
When the conversion $(\vec{\beta},\vec{\lambda})$ to $(\vec{\beta}',\vec{\lambda}')$ by the adiabatic transformation is possible, the value of the extracted work $\tilde{W}$ during the adiabatic transformation satisfies the following equality:
\begin{align}
\tilde{W}=-\Delta \tilde{U}_{S}.
\end{align}
\end{corollary}

We emphasize that Theorem \ref{Tmain} is derived from the assumption that the probability of energy of Gibbs states satisfies the large deviation condition \eqref{LDf1} and \eqref{LDf2}.
We do not impose stronger assumptions, including i.i.d. feature.

In thermodydynamics, from the first law and the second law for adiabatic processes, the other forms of the first law and the second law are derived, e.g. the principle of maximal work and Carnot's theorem and Clausius inequality.
From Theorem \ref{Tmain}, we can derive other forms of the first law and the second law of thermodynamics,  by the same way as in thermodynamics.
In the next subsection, we derive the principle of the first law and the second law for the isothermal processes as examples.

\subsection{Corollary: First Law and Second Law of Thermodynamics for Isothermal Processes}

Now, let us consider the isothermal processes.
We fix an inverse temperature $\beta$ arbitrarily.
Let us take two thermodynamic systems $S$ and $B_{M}$.
The system $S$ is the working body, and the system $B_{M}$ is the heat bath.
Here, $M$ is a real positive number which expresses the ``size'' of  $B_{M}$, and we assume that the specific heat of $B_{M}$ is proportional to $M$, at least when the temperature of $B_{M}$ is close to $1/\beta$.
We firstly define the possibility of the work extraction by isothermal process based on the adiabatic work extraction as follows:$\\$
\textit{isothermal work extraction}:
Let us fix the thermodynamic system $S$, and fix its control parameters $\lambda_{S}$ and $\lambda'_{S}$ arbitrarily.
We also fix an arbitrary real number $\tilde{W}$.
Then, we define ``the isothermal work extraction $(\beta,\lambda_{S})\rightarrow_{\mathrm{is}:\tilde{W}}(\beta,\lambda'_{S})$ is possible'' as follows:$\\$
There exist two functions $\beta'_{M}$ and $\tilde{W}_{M}$ of $M$ satisfying $\lim_{M\rightarrow\infty}\beta'_{M}=\beta$ and $\lim_{M\rightarrow\infty}\tilde{W}_{M}=\tilde{W}$, and 
the adiabatic work extraction $(\beta,\lambda_{S})\times(\beta,\lambda_{B})\rightarrow_{\mathrm{ad}:\tilde{W}_{M}}(\beta,\lambda'_{S})\times(\beta'_{M},\lambda_{B})$ is possible for sufficiently large $M$.

Under the above definition, the following two corollaries hold:
\begin{corollary}[Principle of Maximal Work]\Label{PM}
The isothermal work extraction $(\beta,\lambda)\rightarrow_{\mathrm{is}:\tilde{W}}(\beta,\lambda')$ is possible if and only if the following inequality holds:
\begin{align}
\tilde{W}\le-\Delta \tilde{F}_{S}\Label{pmw}
\end{align}
where 
\begin{align}
\Delta \tilde{F}_{S}&:=\tilde{F}_{S}(\beta,\lambda'_{S})-\tilde{F}_{S}(\beta,\lambda_{S}).
\end{align}
\end{corollary}
\begin{corollary}[First Law of Thermodynamics]\Label{1st}
When the isothermal work extraction $(\beta,\lambda)\rightarrow_{\mathrm{is}:\tilde{W}}(\beta,\lambda')$ is possible, there exists at least one pair of $\tilde{W}_{M}$ and $\beta'_{M}$ satisfying the conditions in the definition of the isothermal work extraction. For any such pair, the limit $\tilde{Q}:=\lim_{M\rightarrow\infty}(\tilde{U}_{B_{M}}(\beta,\lambda_{B})-\tilde{U}_{B_{M}}(\beta'_{M},\lambda_{B}))$ exists, which is independent of $\tilde{W}_{M}$ and $\beta'_{M}$, and satisfies
\begin{align}
\tilde{W}&=\tilde{Q}-\Delta \tilde{U}_{S}.
\end{align}
\end{corollary}
We prove these two corollaries in Appendix \ref{PC1}.

\section{Conclusion}

We have shown that the first law and the second law are derived from an assumption that ``probability distributions of energy in Gibbs states satisfy large deviation'', which is widely accepted as a property of thermodynamic equilibrium states.
As the second law, we have shown that an adiabatic transformation from a set of Gibbs states to another set of Gibbs states is possible if and only if the regularized von Neumann entropy becomes large:
\begin{align}
\Delta \tilde{S}_{S}\ge0.
\end{align}
As the first law, we have shown that the energy loss of the thermodynamic system during the adiabatic transformation is stored in the work storage as ``work,'' in the following meaning;
(i) the energy of the work storage takes certain values macroscopically, in the initial state and the final state.
(ii) the entropy of the work storage in the final state is macroscopically equal to the entropy of the initial state.

As corollaries, our results gives other forms of the first law and the second law.
As an example, we give the maximam work principle $W\le-\Delta F$ and the first law $W=-\Delta U_{S}+Q$ in the isothermal processes.

\textit{Acknowledgments:}
The authors are grateful to Prof. Hiroshi Nagaoka, Prof. Hal Tasaki and Mr. Naoto Shiraishi for helpful comments.
This work was partially supported by JSPS Grant-in-Aid for Scientific Research (C) No. 16K00012 (TO).

\widetext

\appendix
\section{Preliminary: quantum information spectrum method}

In this section, we review the information spectrum method \cite{Han,J-W-O}, which plays a crucial role in the proofs of the main theorems. The information spectrum method enables us to treat coding problems for general information sources and general communication channels, which do not necessarily have the i.i.d. (independent and identically distributed) structure and may have arbitrary correlation. In the information spectrum method, the sup- and inf- spectral entropy rates are two of the most fundamental quantities, and are defined as follows:
\begin{definition}
Let $\widehat\rho=\{\rho_n\}_{n=1}^{\infty}$ be general sequences of quantum states on ${\mathcal H}_n$ whose dimension is $d^{n}$ $(n=1,2,\dots)$.
Then, the sup- and inf- spectral entropy rates $\overline{H}(\widehat\rho)$ and $\underline{H}(\widehat\rho)$ are defined by
\begin{align}
&\overline{H}(\widehat\rho):=\inf\left\{\gamma:\liminf_{n\rightarrow\infty}{\rm Tr}\left[\{\rho_n\geq e^{-n\gamma}I_n\}\rho_n\right]=1\right\},\\
&\underline{H}(\widehat\rho):=\sup\left\{\gamma:\limsup_{n\rightarrow\infty}{\rm Tr}\left[\{\sigma_n\geq e^{-n\gamma}I_n\}\sigma_n\right]=0\right\},\nonumber
\end{align}
Here, $I_n$ is the identity operator on ${\mathcal H}_n$, and $\{\rho_n\geq2^{-n\gamma}I_n\}$ is the projection operator on a subspace of ${\mathcal H}_n$ spanned by the eigenvectors of $\rho_n$ with eigenvalues not smaller than $2^{-n\gamma}$.
\end{definition}

The possibility of asymptotic conversion between two general sequences of quantum states by unital CPTP maps is characterized by using the spectral entropy rate as follows \cite{J-W-O}:

\begin{lemma}\Label{LJWO}
Let $\widehat\rho=\{\rho_n\}_{n=1}^{\infty}$ and $\widehat\sigma=\{\sigma_n\}_{n=1}^{\infty}$
be general sequences of quantum states on ${\mathcal H}_n$ whose dimension is $d^{n}$ $(n=1,2,\dots)$. If $\overline{H}(\widehat\rho)<\underline{H}(\widehat\sigma)$ holds, then there exists a sequence of unital CPTP maps ${\mathcal R}_n$ $(n=1,2,\dots)$ such that
\begin{align}
\lim_{n\rightarrow\infty}\left\|{\mathcal R}_n(\rho_n)-\sigma_n\right\|_1=0.
\end{align}
Moreover, when $\rho_{n}$ and $\sigma_{n}$ are diagonalized by common basis $\{\ket{x}\}^{d^n}_{x=1}$,
there exist a probability $\{q^{(n)}_{j}\}^{d^n}_{j=1}$ and a set of invertible functions $\{f^{(n)}_{j}\}^{d^n}_{j=1}$ from $\{x\}^{d^n}_{x=1}$ to $\{x\}^{d^n}_{x=1}$ satisfying
\begin{align}
\ca{R}_{n}(\rho_{n})=\sum^{d^n}_{j=1}q^{(n)}_{j}\ket{f^{(n)}_{j}(x)}\bra{x}\rho_{n}\ket{x}\bra{f^{(n)}_{j}(x)}.
\end{align}
\end{lemma}
Lemma \ref{LJWO} has been given in \cite{J-W-O}.

When the sup and inf spectral entropies of $\widehat\rho$ satisfies $\overline{H}(\widehat\rho)=\underline{H}(\widehat\rho)=H$ for a real number $H$, we say that  $\{\rho_n\}_{n=1}^{\infty}$ has the entropy spectrum $H$.
Note that because of the large deviation assumptions \eqref{LDf1} and \eqref{LDf2},  the sequence $\{\rho^{(n)}_{\vec{\beta}|\vec{\lambda}}\}$ has the entropy spectrum $\tilde{S}_{S}(\vec{\beta},\vec{\lambda})$.

\section{Proof of Theorem \ref{Tmain}}

Hereafter, we use the abbreviations $\tilde{U}_{S}(\vec{\beta},\vec{\lambda}):=\sum_{k}\tilde{U}_{k}(\beta_{k},\lambda_{k})$, $\tilde{S}_{S}(\vec{\beta},\vec{\lambda}):=\sum_{k}\tilde{S}_{k}(\beta_{k},\lambda_{k})$, $\beta_{\mathrm{max}}:=\max\{\beta_{1},...,\beta_{m}\}$ and $\beta''_{\mathrm{max}}:=\max\{\beta_{1},...,\beta_{m},\beta'_{1},...,\beta'_{m}\}$.

We prove Theorem \ref{Tmain} with using the following lemma:
\begin{lemma}\Label{L1inP}
Let us assume that $\tilde{S}_{S}(\vec{\beta},\vec{\lambda})<\tilde{S}_{S}(\vec{\beta}',\vec{\lambda}')$ holds.
Let us fix an arbitrary positive real number $\delta>0$.
We also fix an arbitrary real positive number $\tilde{U}_{W}$satisfying
\begin{align}
\tilde{U}_{W}&>|\tilde{U}_{S}(\vec{\beta},\vec{\lambda})-\tilde{U}_{S}(\vec{\beta}',\vec{\lambda}')|+8\delta.\Label{Uw}
\end{align}
Then, for an arbitrary  $\delta'>0$ satisfying  $\delta'< \tilde{S}_{S}(\vec{\beta}',\vec{\lambda}')-\tilde{S}_{S}(\vec{\beta},\vec{\lambda})$ and $\delta'<2\beta''_{\mathrm{max}}\delta$, there exist a natural number $N_{\delta'}$ and randomized energy-preserving unitary $\{q^{(n)}_{j,\delta'},V^{(n)}_{j,\delta'}\}^{\infty}_{n=N_{\delta'}}$ whose unitaries $\{V^{(n)}_{j}\}$ satisfy the following three conditions for arbitrary $n\ge N_{\delta'}$:
\begin{align}
\|\sigma^{(n)}_{I,\delta'}-\rho^{(n)}_{\vec{\beta}'|\vec{\lambda}'}\otimes\ket{\vec{\lambda}'}\bra{\vec{\lambda}'}\|&<\delta'\Label{awe1'}\\
\|\sigma^{(n)}_{E_{W},\delta'}-\pi^{(n)}_{\tilde{U}_{W}-\Delta\tilde{U}_{S},\delta}\|&<\left(1+\frac{4}{\beta''_{\mathrm{max}}\delta}\right)\delta'\Label{awe3'}\\
\frac{|S(\sigma^{(n)}_{E_{W},\delta'})-S(\pi^{(n)}_{\tilde{U}_{W},\delta})|}{n}&<\delta',\Label{awe4'}
\end{align}
where  
\begin{align}
\sigma^{(n)}_{IE_{W},\delta'}:=\sum_{j}q^{(n)}_{j,\delta'}V^{(n)}_{j,\delta'}\rho^{(n)}_{\vec{\beta}|\vec{\lambda}}\otimes\ket{\vec{\lambda}}\bra{\vec{\lambda}}\otimes\pi^{(n)}_{\tilde{U}_{W},\delta}V^{(n)\dagger}_{j,\delta'},
\end{align}
and $\sigma^{(n)}_{I,\delta'}:=\Tr_{E_{W}}[\sigma^{(n)}_{IE_{W},\delta'}]$ and $\sigma^{(n)}_{E_{W},\delta'}:=\Tr_{I}[\sigma^{(n)}_{IE_{W},\delta'}]$.
\end{lemma}
Lemma \ref{L1inP} is proved in Appendix C.

Let us prove Theorem \ref{Tmain}.
We firstly show that when $\tilde{S}_{S}(\vec{\beta},\vec{\lambda})<\tilde{S}_{S}(\vec{\beta}',\vec{\lambda}')$ and $\tilde{W}=-\Delta \tilde{U}_{S}$ hold, the adiabatic work extraction $(\beta_{1},\lambda_{1})\times...\times(\beta_{m},\lambda_{m})\rightarrow_{\mathrm{ad}:\tilde{W}}(\beta'_{1},\lambda'_{1})\times...\times(\beta'_{m},\lambda'_{m})$ is possible.
Let us fix an arbitrary $\delta>0$.
We also take the positive number $\tilde{U}_{W}$ satisfying \eqref{Uw}.
We refer to the smallest natural number $k$ satisfying $\frac{1}{k}\le\tilde{S}_{S}(\vec{\beta}',\vec{\lambda}')-\tilde{S}_{S}(\vec{\beta},\vec{\lambda})$ and $\frac{1}{k}\le2\beta''_{\max}\delta$ as $k_{0}$.
For a sequence of natural numbers $\{k\}^{\infty}_{k=k_{0}}$, we define $\delta'_{k}:=\frac{1}{k}$ and take the natural number $N_{\delta'_{k}}$ and the randomized unitary $\{q^{(n)}_{j,\delta'},V^{(n)}_{j,\delta'_{k}}\}^{\infty}_{n=N_{\delta'_{k}}}$ by substituting $\delta'_{k}$ for $\delta'$ in Lemma \ref{L1inP}.
We also define a natural number $\{\tilde{N}_{k}\}^{\infty}_{k=k_{0}}$ as
\begin{align}
\tilde{N}_{k_{0}}&:=\tilde{N}_{\delta'_{k_{0}}},\nonumber\\
\tilde{N}_{k}&:=\max\{\tilde{N}_{\delta'_{k}},\tilde{N}_{\delta'_{k-1}}+1\}.
\end{align}
We also define a randomized shift-invariant and energy preserving unitary $\{q^{(n)}_{j},V^{(n)}_{j}\}^{\infty}_{n=\tilde{N}_{k_{0}}}$ as
\begin{align}
q^{(n)}_{j}&:=q^{(n)}_{j,\delta'_{k}}\enskip(\mbox{for}\enskip \tilde{N}_{k+1}-1\ge n\ge\tilde{N}_{k})\\
V^{(n)}_{j}&:=V^{(n)}_{j,\delta'_{k}}\enskip(\mbox{for}\enskip \tilde{N}_{k+1}-1\ge n\ge\tilde{N}_{k}).
\end{align}
Then,  $\{q^{(n)}_{j},V^{(n)}_{j}\}^{\infty}_{n=\tilde{N}_{k_{0}}}$ satysfies \eqref{awe1}--\eqref{awe4}  because of \eqref{awe1'}--\eqref{awe4'} and $\tilde{W}=-\Delta \tilde{U}_{S}$.
Therefore, when $\tilde{S}_{S}(\vec{\beta},\vec{\lambda})<\tilde{S}_{S}(\vec{\beta}',\vec{\lambda}')$ and $\tilde{W}=-\Delta \tilde{U}_{S}$ hold, the adiabatic work extraction $(\beta_{1},\lambda_{1})\times...\times(\beta_{m},\lambda_{m})\rightarrow_{\mathrm{ad}:\tilde{W}}(\beta'_{1},\lambda'_{1})\times...\times(\beta'_{m},\lambda'_{m})$ is possible.

We secondly show that when $\tilde{S}_{S}(\vec{\beta},\vec{\lambda})=\tilde{S}_{S}(\vec{\beta}',\vec{\lambda}')$ and $\tilde{W}=-\Delta \tilde{U}_{S}$ hold, the adiabatic work extraction $(\beta_{1},\lambda_{1})\times...\times(\beta_{m},\lambda_{m})\rightarrow_{\mathrm{ad}:\tilde{W}}(\beta'_{1},\lambda'_{1})\times...\times(\beta'_{m},\lambda'_{m})$ is possible.
We refer to the smallest natural number $t$ satisfying $\frac{1}{t}<2\beta''_{\max}\delta$ as $t_{0}$, and define a sequence of natural numbers $\{t\}^{\infty}_{t=t_{0}}$.
For each natural number $t$, we define $\vec{\beta}'_{t}:=(\beta'_{1},...,\beta'_{m-1},\beta'_{m}(1-\frac{1}{t})$.
Then, because of $\tilde{S}_{S}(\vec{\beta},\vec{\lambda})<\tilde{S}_{S}(\vec{\beta}',\vec{\lambda}')$ and Lemma \ref{L1inP}, there exist a natural number $N''_{t}$ and a randomized unitary $\{q''^{(n)}_{j,t},V''^{(n)}_{j,t}\}^{\infty}_{n=N''_{t}}$ satisfying the followings:$\\$
\begin{align}
\|\sigma^{(n)}_{I,t}-\rho^{(n)}_{\vec{\beta}'_{t}|\vec{\lambda}'}\otimes\ket{\vec{\lambda}'}\bra{\vec{\lambda}'}\|&<\frac{1}{t}\Label{awe1''}\\
\|\sigma^{(n)}_{E_{W},t}-\pi^{(n)}_{\tilde{U}_{W}-\Delta_{t}\tilde{U}_{S},\delta}\|&<\left(1+\frac{4}{\beta''_{\mathrm{max}}\delta}\right)\frac{1}{t}\Label{awe3''}\\
\frac{|S(\sigma^{(n)}_{E_{W},t})-S(\pi^{(n)}_{\tilde{U}_{W},\delta})|}{n}&<\frac{1}{t}.\Label{awe4''}
\end{align}
where $\Delta_{t}\tilde{U}_{S}:=\tilde{U}_{S}(\vec{\beta}'_{t},\vec{\lambda}')-\tilde{U}_{S}(\vec{\beta},\vec{\lambda})$ and
\begin{align}
\sigma^{(n)}_{IE_{W},t}&:=\sum_{j}q''^{(n)}_{j,t}V''^{(n)}_{j,t}\rho^{(n)}_{\vec{\beta}|\vec{\lambda}}\otimes\ket{\vec{\lambda}}\bra{\vec{\lambda}}\otimes\pi^{(n)}_{\tilde{U}_{W},\delta}V''^{(n)\dagger}_{j,t},\\
\sigma^{(n)}_{I,t}&:=\Tr_{E_{W}}[\sigma^{(n)}_{IE_{W},t}],\enskip\sigma^{(n)}_{E_{W},t}:=\Tr_{I}[\sigma^{(n)}_{IE_{W},t}].
\end{align}
（We can prove the above by substituting $1/t$ for $\delta'$ in Lemma \ref{L1inP}.）
Now, we define
\begin{align}
f(t)&:=\|\rho^{(n)}_{\vec{\beta}'_{t},\vec{\lambda}'}\otimes\ket{\vec{\lambda}'}\bra{\vec{\lambda}'}-\rho^{(n)}_{\vec{\beta}',\vec{\lambda}'}\otimes\ket{\vec{\lambda}'}\bra{\vec{\lambda}'}\|\\
g(t)&:=\pi^{(n)}_{\tilde{U}_{W}-\Delta_{t}\tilde{U}_{S},\delta}.
\end{align}
Because $\tilde{U}$ and $\tilde{S}$ are continuous functions of $\beta$, we obtain $\lim_{t\rightarrow\infty}f(t)=\lim_{t\rightarrow\infty}g(t)=0$ and 
\begin{align}
\|\sigma^{(n)}_{I,t}-\rho^{(n)}_{\vec{\beta}'|\vec{\lambda}'}\otimes\ket{\vec{\lambda}'}\bra{\vec{\lambda}'}\|&<\frac{1}{t}+f(t)\Label{awe1'''}\\
\|\sigma^{(n)}_{E_{W},t}-\pi^{(n)}_{\tilde{U}_{W}-\Delta\tilde{U}_{S},\delta}\|&<\left(1+\frac{4}{\beta''_{\mathrm{max}}\delta}\right)\frac{1}{t}+g(t)\Label{awe3'''}\\
\frac{|S(\sigma^{(n)}_{E_{W},t})-S(\pi^{(n)}_{\tilde{U}_{W},\delta})|}{n}&<\frac{1}{t}.\Label{awe4'''}
\end{align}
Therefore, we define $\{\tilde{N}''_{t}\}^{\infty}_{t=t_{0}}$ as
\begin{align}
\tilde{N}''_{t_{0}}&=N''_{t_{0}}\\
\tilde{N}''_{t}&=\max\{N''_{t},N''_{t-1}+1\},
\end{align}
and define $\{\tilde{q}''^{(n)}_{j,t},\tilde{V}''^{(n)}_{j,t}\}^{\infty}_{n=N''_{t_{0}}}$ as
\begin{align}
\tilde{q}''^{(n)}_{j}&:=q''^{(n)}_{j,t}\enskip(\mbox{for}\enskip \tilde{N}''_{t+1}-1\ge n\ge\tilde{N}''_{t})\\
\tilde{V}''^{(n)}_{j}&:=V''^{(n)}_{j,t}\enskip(\mbox{for}\enskip \tilde{N}''_{t+1}-1\ge n\ge\tilde{N}''_{t}).
\end{align}
Then, $\{\tilde{q}''^{(n)}_{j,t},\tilde{V}''^{(n)}_{j,t}\}^{\infty}_{n=N''_{t_{0}}}$ satisfies \eqref{awe1}--\eqref{awe4} because of \eqref{awe1'''}--\eqref{awe4'''} and $\tilde{W}=\Delta \tilde{U}_{S}$.

Next, we show that when $\tilde{S}_{S}(\vec{\beta},\vec{\lambda})> \tilde{S}_{S}(\vec{\beta}',\vec{\lambda}')$ or $\tilde{W}\ne-\Delta \tilde{U}_{S}$ holds, 
the adiabatic work extraction $(\beta_{1},\lambda_{1})\times...\times(\beta_{m},\lambda_{m})\rightarrow_{\mathrm{ad}:\tilde{W}}(\beta'_{1},\lambda'_{1})\times...\times(\beta'_{m},\lambda'_{m})$ is impossible.
Let us assume that $(\beta_{1},\lambda_{1})\times...\times(\beta_{m},\lambda_{m})\rightarrow_{\mathrm{ad}:\tilde{W}}(\beta'_{1},\lambda'_{1})\times...\times(\beta'_{m},\lambda'_{m})$ would be possible even when $\tilde{S}_{S}(\vec{\beta},\vec{\lambda})> \tilde{S}_{S}(\vec{\beta}',\vec{\lambda}')$ holds.
Then, there exists $\{p^{(n)}_{j},V^{(n)}_{j}\}$ satisfying \eqref{awe1}--\eqref{awe4}.
For the simplicity in expressions, hereafter we use the following description:
\begin{align}
\Lambda^{\rho_{E_{W}}}_{I}(\rho_{I}):=\Tr_{E_{W}}[\sum_{j}p^{(n)}_{j}V^{(n)}_{j}\rho_{I}\otimes\rho_{E_{W}}V^{(n)\dagger}_{j}],\\
\Lambda^{\rho_{I}}_{E_{W}}(\rho_{E_{W}}):=\Tr_{I}[\sum_{j}p^{(n)}_{j}V^{(n)}_{j}\rho_{I}\otimes\rho_{E_{W}}V^{(n)\dagger}_{j}].
\end{align}
where $\rho_{I}$ and $\rho_{E_{W}}$ are arbitrary states of $I$ and $E_{W}$, respectively.

Let us derive a contradiction from the existence of $\{p^{(n)}_{j},V^{(n)}_{j}\}$.
Firstly, we define the region $1_{Con}$ as the subspace of $\ca{H}^{(n)}_{I}$ which is spanned by the energy eigenvecters of $H^{(n)}_{I}$ whose eigenvalues $E^{(n)}$ satisfy
\begin{align}
n(\tilde{U}_{S}(\vec{\beta},\vec{\lambda})-\epsilon)\le E^{(n)}\le n(\tilde{U}_{S}(\vec{\beta},\vec{\lambda})+\epsilon)\Label{Condef}
\end{align}
where ``Con'' is abbreviation of ``Converse part.''
Next, we take the following approximate state of $\rho^{(n)}_{\vec{\beta}|\vec{\lambda}}\otimes\ket{\vec{\lambda}}\bra{\vec{\lambda}}$:
\begin{align}
\tilde{\rho}^{(n)Con}_{\vec{\beta}|\vec{\lambda}}\otimes\ket{\vec{\lambda}}\bra{\vec{\lambda}}&:=\frac{\Pi^{(n)}_{\mathrm{region1_{Con}}}  \rho^{(n)}_{\vec{\beta}|\vec{\lambda}}\otimes\ket{\vec{\lambda}}\bra{\vec{\lambda}}  \Pi^{(n)}_{\mathrm{region1_{Con}}}    }{\Tr[\Pi^{(n)}_{\mathrm{region1_{Con}}}  \rho^{(n)}_{\vec{\beta}|\vec{\lambda}} \otimes\ket{\vec{\lambda}}\bra{\vec{\lambda}} ] }
\end{align}
where $\Pi^{(n)}_{\mathrm{region1_{Con}}}$ is the projection to the region $1_{Con}$.

Next, we refer to the support of $\Lambda^{\pi^{(n)}_{\tilde{U}_{W},\delta}}_{I}(\tilde{\rho}^{(n)Con}_{\vec{\beta}|\vec{\lambda}}\otimes\ket{\vec{\lambda}}\bra{\vec{\lambda}})$ as the region $2'_{Con}$.
We also define the region $2_{Con}$ as the subspace of $\ca{H}^{(n)}_{I}$ which is spanned by the energy eigenvecters of $H^{(n)}_{I}$ whose eigenvalues $E^{(n)}$ satisfy 
\begin{align}
n(\tilde{U}_{S}(\vec{\beta}',\vec{\lambda}')-\epsilon')\le E^{(n)}\le n(\tilde{U}_{S}(\vec{\beta}',\vec{\lambda}')+\epsilon')\Label{A26}.
\end{align}
Because the Hamiltonian $H^{(n)}_{E_{W}}$ has a ground energy level and each $V^{(n)}_{j}$ is energy-preserving, if we take $\epsilon'>0$ enough large, the region $2_{Con}$ includes the region $2'_{Con}$.
Then, we define an approximate state of $\rho^{(n)}_{\vec{\beta}'|\vec{\lambda}'}\otimes\ket{\vec{\lambda}'}\bra{\vec{\lambda}'}$ as follows:
\begin{align}
\tilde{\rho}^{(n)Con}_{\vec{\beta}'|\vec{\lambda}'}\otimes\ket{\vec{\lambda}'}\bra{\vec{\lambda}'}&:=\frac{\Pi^{(n)}_{\mathrm{region 2_{Con}}}  \rho^{(n)}_{\vec{\beta}'|\vec{\lambda}'}\otimes\ket{\vec{\lambda}'}\bra{\vec{\lambda}'} \Pi^{(n)}_{\mathrm{region2_{Con}}}       }{\Tr[\Pi^{(n)}_{\mathrm{region2_{Con}}}  \rho^{(n)}_{\vec{\beta}'|\vec{\lambda}'}\otimes\ket{\vec{\lambda}'}\bra{\vec{\lambda}'}  ] }.
\end{align}
where $\Pi^{(n)}_{\mathrm{region2_{Con}}}$ is the projection to the region $2_{Con}$.

Because of the large deviation assumptions \eqref{LDf1} and \eqref{LDf2}, there exists a real positive number $\alpha$, the inequalities
\begin{align}
\Tr[\Pi^{(n)}_{\mathrm{region1_{Con}}} \rho^{(n)}_{\vec{\beta}|\vec{\lambda}}\otimes\ket{\vec{\lambda}}\bra{\vec{\lambda}} ]&\ge1-e^{-n\alpha}\\
\Tr[\Pi^{(n)}_{\mathrm{region2_{Con}}} \rho^{(n)}_{\vec{\beta}'|\vec{\lambda}'}\otimes\ket{\vec{\lambda}'}\bra{\vec{\lambda}'}]&\ge1-e^{-n\alpha}\Label{A31}
\end{align}
hold for sufficiently large $n$.
Therefore, because of the gentle measurement lemma \cite{gentle}, the following inequalities hold for the sufficiently large $n$:
\begin{align}
\|\rho^{(n)}_{\vec{\beta}|\vec{\lambda}}\otimes\ket{\vec{\lambda}}\bra{\vec{\lambda}}-\tilde{\rho}^{(n)Con}_{\vec{\beta}|\vec{\lambda}}\otimes\ket{\vec{\lambda}}\bra{\vec{\lambda}}\|\le e^{-n\alpha/2}\Label{app1Con}\\
\|\rho^{(n)}_{\vec{\beta}'|\vec{\lambda}'}\otimes\ket{\vec{\lambda}'}\bra{\vec{\lambda}'}-\tilde{\rho}^{(n)Con}_{\vec{\beta}'|\vec{\lambda}'}\otimes\ket{\vec{\lambda}'}\bra{\vec{\lambda}'}\|\le e^{-n\alpha/2}\Label{app1'Con}
\end{align}
Also, for arbitrary $\epsilon>0$, $\vec{\beta}$ and $\vec{\lambda}$, the following relations hold:
\begin{align}
\lim_{n\rightarrow\infty}\frac{S(\tilde{\rho}^{(n)Con}_{\vec{\beta}|\vec{\lambda}}\otimes\ket{\vec{\lambda}}\bra{\vec{\lambda}})}{n}&=\tilde{S}_{S}(\vec{\beta},\vec{\lambda}),\Label{entropymain}\\
\lim_{n\rightarrow\infty}\frac{(1-r^{Con}_{\mathrm{main}})S(\tilde{\rho}^{(n)Con\lnot}_{\vec{\beta}|\vec{\lambda}}\otimes\ket{\vec{\lambda}}\bra{\vec{\lambda}}))}{n}&=0,\Label{entropysub}
\end{align}
where 
\begin{align}
r^{Con}_{\mathrm{main}}&:=\Tr[\Pi^{(n)}_{\mathrm{region1_{Con}}}  \rho^{(n)}_{\vec{\beta}|\vec{\lambda}} \otimes\ket{\vec{\lambda}}\bra{\vec{\lambda}} ] \\
\tilde{\rho}^{(n)Con\lnot}_{\vec{\beta}|\vec{\lambda}}\otimes\ket{\vec{\lambda}}\bra{\vec{\lambda}}&:=\frac{(1-\Pi^{(n)}_{\mathrm{region1_{Con}}})  \rho^{(n)}_{\vec{\beta}|\vec{\lambda}}\otimes\ket{\vec{\lambda}}\bra{\vec{\lambda}} (1- \Pi^{(n)}_{\mathrm{region1_{Con}}} )   }{\Tr[(1-\Pi^{(n)}_{\mathrm{region1_{Con}}} ) \rho^{(n)}_{\vec{\beta}|\vec{\lambda}}\otimes\ket{\vec{\lambda}}\bra{\vec{\lambda}}  ] }
\end{align}
(\textit{Proof of \eqref{entropymain} and \eqref{entropysub}:}
We firstly show that \eqref{entropymain}.
Note that the sequences $\{\rho^{(n)}_{\vec{\beta}|\vec{\lambda}}\}$ and $\{\tilde{\rho}^{(n)}_{\vec{\beta}|\vec{\lambda}}\}$ have the same entropy spectrum $\tilde{S}_{S}(\vec{\beta},\vec{\lambda})$, and that the dimension of region $1_{Con}$ is lower than $e^{n(\beta_{\max}\tilde{U}_{S}(\vec{\beta},\vec{\lambda})-\sum_{k}\tilde{F}_{k}(\beta_{k},\lambda_{k})+\alpha)}$ for sufficiently large $n$.
When the dimension of $\ca{H}_{n}$ is lower than $D^{n}$ for finite $D$ and when $\{\rho_{n}\}$ has the entropy spectrum $H$, the entropy rate of  $\{\rho_{n}\}$ is equal to $H$.  
Therefore, the entropy rate of $\{\tilde{\rho}^{(n)}_{\vec{\beta}|\vec{\lambda}}\}$ is equal to $\tilde{S}_{S}(\vec{\beta},\vec{\lambda})$, and thus \eqref{entropymain} holds.

Next, we show \eqref{entropysub}.
The relation \eqref{entropysub} are easily derived from the following inequality:
\begin{align}
S(\rho^{(n)}_{\vec{\beta}|\vec{\lambda}}\otimes\ket{\vec{\lambda}}\bra{\vec{\lambda}})&=S(\rho^{(n)}_{\vec{\beta}|\vec{\lambda}})
=S(r^{Con}_{\mathrm{main}}\tilde{\rho}^{(n)Con}_{\vec{\beta}|\vec{\lambda}}+(1-r^{Con}_{\mathrm{main}})\tilde{\rho}^{(n)Con\lnot}_{\vec{\beta}|\vec{\lambda}})\nonumber\\
&\ge r^{Con}_{\mathrm{main}}S(\tilde{\rho}^{(n)Con}_{\vec{\beta}|\vec{\lambda}})+(1-r^{Con}_{\mathrm{main}})S(\tilde{\rho}^{(n)Con\lnot}_{\vec{\beta}|\vec{\lambda}})\nonumber\\
&\stackrel{(b)}{\ge}r^{Con}_{\mathrm{main}}(n\tilde{S}_{S}(\vec{\beta},\vec{\lambda})+o(n))+(1-r^{Con}_{\mathrm{main}})S(\tilde{\rho}^{(n)Con\lnot}_{\vec{\beta}|\vec{\lambda}})
\end{align}
where $(b)$ is given by \eqref{entropymain}. \textit{Proof end})

Because of \eqref{awe1}, we obtain
\begin{align}
\lim_{n\rightarrow\infty}\|\Lambda^{\pi^{(n)}_{\tilde{U}_{W},\delta}}_{I}(\rho^{(n)}_{\vec{\beta}|\vec{\lambda}}\otimes\ket{\vec{\lambda}}\bra{\vec{\lambda}})-\rho^{(n)}_{\vec{\beta}'|\vec{\lambda}'}\otimes\ket{\vec{\lambda}'}\bra{\vec{\lambda}'}\|=0.
\end{align}
Because of \eqref{app1Con}, we obtain
\begin{align}
\lim_{n\rightarrow\infty}\|\Lambda^{\pi^{(n)}_{\tilde{U}_{W},\delta}}_{I}(\rho^{(n)}_{\vec{\beta}|\vec{\lambda}}\otimes\ket{\vec{\lambda}}\bra{\vec{\lambda}})
-\Lambda^{\pi^{(n)}_{\tilde{U}_{W},\delta}}_{I}(\tilde{\rho}^{(n)Con}_{\vec{\beta}|\vec{\lambda}}\otimes\ket{\vec{\lambda}}\bra{\vec{\lambda}})\|=0.
\end{align}
Because of \eqref{app1'Con}, we obtain
\begin{align}
\lim_{n\rightarrow\infty}\|\rho^{(n)}_{\vec{\beta}'|\vec{\lambda}'}\otimes\ket{\vec{\lambda}'}\bra{\vec{\lambda}'}
-
\tilde{\rho}^{(n)Con}_{\vec{\beta}'|\vec{\lambda}'}\otimes\ket{\vec{\lambda}'}\bra{\vec{\lambda}'}
\|=0.
\end{align}
Therefore,
\begin{align}
\lim_{n\rightarrow\infty}\|\Lambda^{\pi^{(n)}_{\tilde{U}_{W},\delta}}_{I}(\tilde{\rho}^{(n)Con}_{\vec{\beta}|\vec{\lambda}}\otimes\ket{\vec{\lambda}}\bra{\vec{\lambda}})-
\tilde{\rho}^{(n)con}_{\vec{\beta}'|\vec{\lambda}'}\otimes\ket{\vec{\lambda}'}\bra{\vec{\lambda}'}\|=0.
\end{align}

Because of \eqref{A26},  the dimension of the region $2_{Con}$ is lower than $e^{n(\beta'_{\max}\tilde{U}_{S}(\vec{\beta}',\vec{\lambda}')-\sum_{k}\tilde{F}_{k}(\beta'_{k},\lambda'_{k})+\gamma)}$ for sufficiently large $n$, where $\gamma$ is a proper positive number.
Therefore, we obtain the following from the Fannes inequality:
\begin{align}
\lim_{n\rightarrow\infty}\frac{1}{n}(S(\Lambda^{\pi^{(n)}_{\tilde{U}_{W},\delta}}_{I}(\tilde{\rho}^{(n)Con}_{\vec{\beta}|\vec{\lambda}}\otimes\ket{\vec{\lambda}}\bra{\vec{\lambda}}))-S(\tilde{\rho}^{(n)Con}_{\vec{\beta}'|\vec{\lambda}'}\otimes\ket{\vec{\lambda}'}\bra{\vec{\lambda}'})=0
\end{align}
Therefore, because of \eqref{entropymain},
\begin{align}
\lim_{n\rightarrow\infty}\frac{1}{n}S(\Lambda^{\pi^{(n)}_{\tilde{U}_{W},\delta}}_{I}(\tilde{\rho}^{(n)Con}_{\vec{\beta}|\vec{\lambda}}\otimes\ket{\vec{\lambda}}\bra{\vec{\lambda}}))=\tilde{S}_{S}(\vec{\beta}',\vec{\lambda}').\Label{A42}
\end{align}

Note that $\Lambda^{\rho^{(n)}_{\vec{\beta}|\vec{\lambda}}\otimes\ket{\vec{\lambda}}\bra{\vec{\lambda}}}_{E_{W}}(\pi_{\tilde{U}_{W},\delta})=r^{Con}_{\mathrm{main}}\Lambda^{\tilde{\rho}^{(n)Con}_{\vec{\beta}}\otimes\ket{\vec{\lambda}}\bra{\vec{\lambda}}}_{E_{W}}(\pi_{\tilde{U}_{W},\delta})+(1-r^{Con}_{\mathrm{main}})\Lambda^{\tilde{\rho}^{(n)\lnot Con}_{\vec{\beta}}\otimes\ket{\vec{\lambda}}\bra{\vec{\lambda}}}_{E_{W}}(\pi_{\tilde{U}_{W},\delta})$.
Therefore, 
\begin{align}
S(\Lambda^{\rho^{(n)}_{\vec{\beta}|\vec{\lambda}}\otimes\ket{\vec{\lambda}}\bra{\vec{\lambda}}}_{E_{W}}(\pi_{\tilde{U}_{W},\delta}))&\ge r^{Con}_{\mathrm{main}}S(\Lambda^{\tilde{\rho}^{(n)Con}_{\vec{\beta}}\otimes\ket{\vec{\lambda}}\bra{\vec{\lambda}}}_{E_{W}}(\pi_{\tilde{U}_{W},\delta}))+(1-r^{Con}_{\mathrm{main}})S(\Lambda^{\tilde{\rho}^{(n)\lnot Con}_{\vec{\beta}}\otimes\ket{\vec{\lambda}}\bra{\vec{\lambda}}}_{E_{W}}(\pi_{\tilde{U}_{W},\delta}))\nonumber\\
&\ge  r^{Con}_{\mathrm{main}}S(\Lambda^{\tilde{\rho}^{(n)Con}_{\vec{\beta}}\otimes\ket{\vec{\lambda}}\bra{\vec{\lambda}}}_{E_{W}}(\pi_{\tilde{U}_{W},\delta}))\Label{A45}.
\end{align}

Now, let us derive the contradiction.
Note that for $\rho'_{A}:=\Tr^{(n)}_{B}[U_{AB}\rho_{A}\otimes\rho_{B}U^{\dagger}_{AB}]$ and $\rho'_{B}:=\Tr^{(n)}_{A}[U_{AB}\rho_{A}\otimes\rho_{B}U^{\dagger}_{AB}]$,
$S(\rho_{A})+S(\rho_{B})\le S(\rho'_{A})+S(\rho'_{B})$ holds.
By substituting $I$ and $E_{W}$ for $A$ and $B$, we obtain
\begin{align}
S(\Lambda^{\pi^{(n)}_{\tilde{U}_{W},\delta}}_{I}(\tilde{\rho}^{(n)}_{\vec{\beta}|\vec{\lambda}}\otimes\ket{\vec{\lambda}}\bra{\vec{\lambda}}))
+S(\Lambda^{\tilde{\rho}^{(n)Con}_{\vec{\beta}}\otimes\ket{\vec{\lambda}}\bra{\vec{\lambda}}}_{E_{W}}(\pi_{\tilde{U}_{W},\delta}))
\ge
S(\tilde{\rho}^{(n)}_{\vec{\beta}|\vec{\lambda}}\otimes\ket{\vec{\lambda}}\bra{\vec{\lambda}}))
+S(\pi_{\tilde{U}_{W},\delta})
\end{align}
Because of \eqref{A45}, we obtain
\begin{align}
S(\Lambda^{\pi^{(n)}_{\tilde{U}_{W},\delta}}_{I}(\tilde{\rho}^{(n)}_{\vec{\beta}|\vec{\lambda}}\otimes\ket{\vec{\lambda}}\bra{\vec{\lambda}}))
+\frac{1}{r^{Con}_{\mathrm{main}}}S(\Lambda^{\rho^{(n)}_{\vec{\beta}}\otimes\ket{\vec{\lambda}}\bra{\vec{\lambda}}}_{E_{W}}(\pi_{\tilde{U}_{W},\delta}))
\ge
S(\tilde{\rho}^{(n)}_{\vec{\beta}|\vec{\lambda}}\otimes\ket{\vec{\lambda}}\bra{\vec{\lambda}}))
+S(\pi_{\tilde{U}_{W},\delta})
\end{align}
Because of \eqref{LDf1} and \eqref{LDf2}, $r^{Con}_{\mathrm{main}}=1-o(e^{-n\alpha})$.
Therefore, we obtain the following from \eqref{awe4}:
\begin{align}
\liminf_{n\rightarrow\infty}\frac{1}{n}(S(\Lambda^{\pi^{(n)}_{\tilde{U}_{W},\delta}}_{I}(\tilde{\rho}^{(n)Con}_{\vec{\beta}|\vec{\lambda}}\otimes\ket{\vec{\lambda}}\bra{\vec{\lambda}}))-S(\tilde{\rho}^{(n)Con}_{\vec{\beta}|\vec{\lambda}}\otimes\ket{\vec{\lambda}}\bra{\vec{\lambda}}))\ge0.\Label{jyousiki}
\end{align}
Because of \eqref{entropymain}, $S(\tilde{\rho}^{(n)Con}_{\vec{\beta}|\vec{\lambda}}\otimes\ket{\vec{\lambda}}\bra{\vec{\lambda}})/n$ goes to $\tilde{S}_{S}(\vec{\beta},\vec{\lambda})$ at the limit of $n\rightarrow\infty$.
We have assumed that $\tilde{S}_{S}(\vec{\beta},\vec{\lambda})> \tilde{S}_{S}(\vec{\beta}',\vec{\lambda}')$.
Therefore \eqref{jyousiki} implies that $S(\Lambda^{\pi^{(n)}_{\tilde{U}_{W},\delta}}_{I}(\tilde{\rho}^{(n)Con}_{\vec{\beta}|\vec{\lambda}}\otimes\ket{\vec{\lambda}}\bra{\vec{\lambda}}))/n$ cannot converge to $\tilde{S}_{S}(\vec{\beta}',\vec{\lambda}')$ at the limit of $n\rightarrow\infty$. This contradicts to \eqref{A42}.
Therefore,  when $\tilde{S}_{S}(\vec{\beta},\vec{\lambda})> \tilde{S}_{S}(\vec{\beta}',\vec{\lambda}')$ holds, 
the adiabatic work extraction $(\beta_{1},\lambda_{1})\times...\times(\beta_{m},\lambda_{m})\rightarrow_{\mathrm{ad}:\tilde{W}}(\beta'_{1},\lambda'_{1})\times...\times(\beta'_{m},\lambda'_{m})$ is impossible.

Next, we show that  when $\tilde{S}_{S}(\vec{\beta},\vec{\lambda})\le \tilde{S}_{S}(\vec{\beta}',\vec{\lambda}')$ and $\tilde{W}\ne-\Delta \tilde{U}_{S}$ hold, 
the adiabatic work extraction $(\beta_{1},\lambda_{1})\times...\times(\beta_{m},\lambda_{m})\rightarrow_{\mathrm{ad}:\tilde{W}}(\beta'_{1},\lambda'_{1})\times...\times(\beta'_{m},\lambda'_{m})$ is impossible.
Let us assume that $(\beta_{1},\lambda_{1})\times...\times(\beta_{m},\lambda_{m})\rightarrow_{\mathrm{ad}:\tilde{W}}(\beta'_{1},\lambda'_{1})\times...\times(\beta'_{m},\lambda'_{m})$ would be possible even when $\tilde{S}_{S}(\vec{\beta},\vec{\lambda})\le\tilde{S}_{S}(\vec{\beta}',\vec{\lambda}')$ and $\tilde{W}\ne-\Delta \tilde{U}_{S}$ hold.
Let us take a real number $\delta$ such that $0<\delta<|-\Delta\tilde{U}_{S}-\tilde{W}|/10$.
Then, there exist $\tilde{U}_{W}$, $h_{0}$ and $\{p^{(n)}_{j},V^{(n)}_{j}\}$ satisfying \eqref{awe1}--\eqref{awe4}.
Let us derive a contradiction from the existence of $\{p^{(n)}_{j},V^{(n)}_{j}\}$.

The assumptions \eqref{LDf1} and \eqref{LDf2} imply the following:
\begin{align}
\lim_{n\rightarrow\infty}\Tr[\Pi^{(n)I}_{\tilde{U}_{S}(\vec{\beta}',\vec{\lambda}'),\delta}\rho^{(n)}_{\vec{\beta}'|\vec{\lambda}'}\otimes\ket{\vec{\lambda}'}\bra{\vec{\lambda}'}]&=1,\Label{11.24.2}\\
\lim_{n\rightarrow\infty}\Tr[\Pi^{(n)I}_{\tilde{U}_{S}(\vec{\beta},\vec{\lambda}),\delta}\rho^{(n)}_{\vec{\beta}|\vec{\lambda}}\otimes\ket{\vec{\lambda}}\bra{\vec{\lambda}}]&=1.\Label{11.24.4}
\end{align}
where $\Pi^{(n)I}_{X,Y}$ is the proposition to the subspace of $\ca{H}^{(n)}_{I}$ spanned by the energy eigenvectors whose eigenvalues $E^{(n)}$ satisfy
\begin{align}
n(X-Y)\le E^{(n)}\le n(X+Y).\Label{A47}
\end{align}
Hereafter, we also use $\Pi^{(n)E_{W}}_{X,Y}$ and $\Pi^{(n)IE_{W}}_{X,Y}$ as the propositions to the subspaces of $\ca{H}^{(n)}_{E_{W}}$ and $\ca{H}^{(n)}_{IE_{W}}$ spanned by the energy eigenvectors whose eigenvalues $E^{(n)}$ satisfy \eqref{A47}, respectively.
From \eqref{awe1} and \eqref{11.24.2}, we obtain
\begin{align}
\lim_{n\rightarrow\infty}\Tr[\Pi^{(n)I}_{\tilde{U}_{S}(\vec{\beta}',\vec{\lambda}'),\delta}\sigma^{(n)}_{I}]=1.\Label{11.24.3}
\end{align}

Because of \eqref{11.24.4},
\begin{align}
\lim_{n\rightarrow\infty}\Tr[\Pi^{(n)IE_{W}}_{\tilde{U}_{S}(\vec{\beta},\vec{\lambda})+\tilde{U}_{W},2\delta}\rho^{(n)}_{\vec{\beta}|\vec{\lambda}}\otimes\ket{\vec{\lambda}}\bra{\vec{\lambda}}\otimes\pi^{(n)}_{\tilde{U}_{W},\delta}]=1.\Label{11.24.5}
\end{align}
Because each $V^{(n)}_{j}$ is an energy preserving unitary, \eqref{11.24.5} implies the following:
\begin{align}
\lim_{n\rightarrow\infty}\Tr[\Pi^{(n)IE_{W}}_{\tilde{U}_{S}(\vec{\beta},\vec{\lambda})+\tilde{U}_{W},2\delta}\sigma^{(n)}_{IE_{W}}]=1.\Label{11.24.6}
\end{align}
Because of the gentle measurement lemma \cite{gentle} and \eqref{11.24.6}, we derive
\begin{align}
\lim_{n\rightarrow\infty}\|\Pi^{(n)IE_{W}}_{\tilde{U}_{S}(\vec{\beta},\vec{\lambda})+\tilde{U}_{W},2\delta}\sigma^{(n)}_{IE_{W}}\Pi^{(n)IE_{W}}_{\tilde{U}_{S}(\vec{\beta},\vec{\lambda})+\tilde{U}_{W},2\delta}-\sigma^{(n)}_{IE_{W}}\|=0\Label{3.25.2}
\end{align}
Similarly, because of the gentle measurement lemma \cite{gentle} and \eqref{11.24.3}, we derive
\begin{align}
\lim_{n\rightarrow\infty}\|\Pi^{(n)I}_{\tilde{U}_{S}(\vec{\beta}',\vec{\lambda}'),\delta}\sigma^{(n)}_{IW}\Pi^{(n)I}_{\tilde{U}_{S}(\vec{\beta}',\vec{\lambda}'),\delta}-\sigma^{(n)}_{IW}\|=0.\Label{3.25.3}
\end{align}
Note that 
\begin{align}
\lim_{n\rightarrow\infty}\|\rho_{n}-\sigma_{n}\|=0\enskip\Rightarrow\enskip\lim_{n\rightarrow\infty}\|P_{n}\rho_{n}P_{n}-P_{n}\sigma_{n}P_{n}\|=0\Label{3.25.4}
\end{align}
for any projection $P_{n}$. (\textit{Proof}: Let us define a CPTP ${\cal E}_{n}(\rho):=P_{n}\rho P_{n}+(1-P_{n})\rho(1-P_{n})$. 
Then, $\|\rho_{n}-\sigma_{n}\|\ge\|\ca{E}_{n}(\rho_{n})-\ca{E}_{n}(\sigma_{n})\|\ge\|P_{n}\rho_{n}P_{n}-P_{n}\sigma_{n}P_{n}\|$.)
Because of \eqref{3.25.2} and \eqref{3.25.4},
\begin{align}
\lim_{n\rightarrow\infty}
\|
\Pi^{(n)I}_{\tilde{U}_{S}(\vec{\beta}',\vec{\lambda}'),\delta}\Pi^{(n)IE_{W}}_{\tilde{U}_{S}(\vec{\beta},\vec{\lambda})+\tilde{U}_{W},2\delta}\sigma^{(n)}_{IE_{W}}\Pi^{(n)IE_{W}}_{\tilde{U}_{S}(\vec{\beta},\vec{\lambda})+\tilde{U}_{W},2\delta}\Pi^{(n)I}_{\tilde{U}_{S}(\vec{\beta}',\vec{\lambda}'),\delta}
-
\Pi^{(n)I}_{\tilde{U}_{S}(\vec{\beta}',\vec{\lambda}'),\delta}\sigma^{(n)}_{IW}\Pi^{(n)I}_{\tilde{U}_{S}(\vec{\beta}',\vec{\lambda}'),\delta}
\|=0\Label{3.25.5}
\end{align}
Because of \eqref{3.25.3} and \eqref{3.25.5},
\begin{align}
\lim_{n\rightarrow\infty}
\|
\Pi^{(n)I}_{\tilde{U}_{S}(\vec{\beta}',\vec{\lambda}'),\delta}\Pi^{(n)IE_{W}}_{\tilde{U}_{S}(\vec{\beta},\vec{\lambda})+\tilde{U}_{W},2\delta}\sigma^{(n)}_{IE_{W}}\Pi^{(n)IE_{W}}_{\tilde{U}_{S}(\vec{\beta},\vec{\lambda})+\tilde{U}_{W},2\delta}\Pi^{(n)I}_{\tilde{U}_{S}(\vec{\beta}',\vec{\lambda}'),\delta}
-
\sigma^{(n)}_{IW}
\|=0\Label{3.25.6}
\end{align}
Because the support of $\Pi^{(n)E_{W}}_{-\Delta\tilde{U}_{S}+\tilde{U}_{W},4\delta}\otimes\hat{1}_{I}$ includes the support of $(\Pi^{(n)I}_{\tilde{U}_{S}(\vec{\beta}',\vec{\lambda}'),\delta}\otimes\hat{1}_{E_{W}})\Pi^{(n)IE_{W}}_{\tilde{U}_{S}(\vec{\beta},\vec{\lambda})+\tilde{U}_{W},2\delta}$,
\begin{align}
\Pi^{(n)E_{W}}_{-\Delta\tilde{U}_{S}+\tilde{U}_{W},4\delta}
\Pi^{(n)I}_{\tilde{U}_{S}(\vec{\beta}',\vec{\lambda}'),\delta}\Pi^{(n)IE_{W}}_{\tilde{U}_{S}(\vec{\beta},\vec{\lambda})+\tilde{U}_{W},2\delta}\sigma^{(n)}_{IE_{W}}\Pi^{(n)IE_{W}}_{\tilde{U}_{S}(\vec{\beta},\vec{\lambda})+\tilde{U}_{W},2\delta}\Pi^{(n)I}_{\tilde{U}_{S}(\vec{\beta}',\vec{\lambda}'),\delta}
\Pi^{(n)E_{W}}_{-\Delta\tilde{U}_{S}+\tilde{U}_{W},4\delta}\nonumber\\
=
\Pi^{(n)I}_{\tilde{U}_{S}(\vec{\beta}',\vec{\lambda}'),\delta}\Pi^{(n)IE_{W}}_{\tilde{U}_{S}(\vec{\beta},\vec{\lambda})+\tilde{U}_{W},2\delta}\sigma^{(n)}_{IE_{W}}\Pi^{(n)IE_{W}}_{\tilde{U}_{S}(\vec{\beta},\vec{\lambda})+\tilde{U}_{W},2\delta}\Pi^{(n)I}_{\tilde{U}_{S}(\vec{\beta}',\vec{\lambda}'),\delta}
\end{align}
Therefore,
\begin{align}
\lim_{n\rightarrow\infty}
\|
\Pi^{(n)E_{W}}_{-\Delta\tilde{U}_{S}+\tilde{U}_{W},4\delta}
\Pi^{(n)I}_{\tilde{U}_{S}(\vec{\beta}',\vec{\lambda}'),\delta}\Pi^{(n)IE_{W}}_{\tilde{U}_{S}(\vec{\beta},\vec{\lambda})+\tilde{U}_{W},2\delta}\sigma^{(n)}_{IE_{W}}\Pi^{(n)IE_{W}}_{\tilde{U}_{S}(\vec{\beta},\vec{\lambda})+\tilde{U}_{W},2\delta}\Pi^{(n)I}_{\tilde{U}_{S}(\vec{\beta}',\vec{\lambda}'),\delta}
\Pi^{(n)E_{W}}_{-\Delta\tilde{U}_{S}+\tilde{U}_{W},4\delta}
-
\sigma^{(n)}_{IW}
\|=0\Label{3.25.7}
\end{align}
Because \eqref{3.25.4} and \eqref{3.25.6},
\begin{align}
\lim_{n\rightarrow\infty}
\|
\Pi^{(n)E_{W}}_{-\Delta\tilde{U}_{S}+\tilde{U}_{W},4\delta}
\Pi^{(n)I}_{\tilde{U}_{S}(\vec{\beta}',\vec{\lambda}'),\delta}\Pi^{(n)IE_{W}}_{\tilde{U}_{S}(\vec{\beta},\vec{\lambda})+\tilde{U}_{W},2\delta}\sigma^{(n)}_{IE_{W}}\Pi^{(n)IE_{W}}_{\tilde{U}_{S}(\vec{\beta},\vec{\lambda})+\tilde{U}_{W},2\delta}\Pi^{(n)I}_{\tilde{U}_{S}(\vec{\beta}',\vec{\lambda}'),\delta}
\Pi^{(n)E_{W}}_{-\Delta\tilde{U}_{S}+\tilde{U}_{W},4\delta}\nonumber\\
-
\Pi^{(n)E_{W}}_{-\Delta\tilde{U}_{S}+\tilde{U}_{W},4\delta}
\sigma^{(n)}_{IW}
\Pi^{(n)E_{W}}_{-\Delta\tilde{U}_{S}+\tilde{U}_{W},4\delta}
\|=0\Label{3.25.8}
\end{align}
Because of \eqref{3.25.7} and \eqref{3.25.8},
\begin{align}
\lim_{n\rightarrow\infty}
\|
\Pi^{(n)E_{W}}_{-\Delta\tilde{U}_{S}+\tilde{U}_{W},4\delta}
\sigma^{(n)}_{IW}
\Pi^{(n)E_{W}}_{-\Delta\tilde{U}_{S}+\tilde{U}_{W},4\delta}
-
\sigma^{(n)}_{IW}
\|=0.
\end{align}
Therefore,
\begin{align}
\lim_{n\rightarrow\infty}\Tr[\Pi^{(n)E_{W}}_{-\Delta\tilde{U}_{S}+\tilde{U}_{W},4\delta}\sigma^{(n)}_{IW}]=1\Label{3.25.9}
\end{align}
On the other hand, because of \eqref{awe3},
\begin{align}
\lim_{n\rightarrow\infty}\Tr[\Pi^{(n)E_{W}}_{\tilde{U}_{W}+\tilde{W},\delta}\sigma^{(n)}_{E_{W}}]=1.\Label{3.25.1}
\end{align}
Becasuse of $\delta<|-\Delta\tilde{U}_{S}-\tilde{W}|/10$ and \eqref{3.25.1},
\begin{align}
\lim_{n\rightarrow\infty}\Tr[\Pi^{(n)E_{W}}_{-\Delta\tilde{U}_{S}+\tilde{U}_{W},4\delta}\sigma^{(n)}_{IW}]=0\Label{3.25.10}
\end{align}
The equalities \eqref{3.25.9} and \eqref{3.25.10} contradict each others.
Therefore, when $\tilde{S}_{S}(\vec{\beta},\vec{\lambda})\le \tilde{S}_{S}(\vec{\beta}',\vec{\lambda}')$ and $\tilde{W}\ne-\Delta \tilde{U}_{S}$ hold, 
the adiabatic work extraction $(\beta_{1},\lambda_{1})\times...\times(\beta_{m},\lambda_{m})\rightarrow_{\mathrm{ad}:\tilde{W}}(\beta'_{1},\lambda'_{1})\times...\times(\beta'_{m},\lambda'_{m})$ is impossible.

\section{Proof of Lemma \ref{L1inP}}

We prove Lemma \ref{L1inP} with using Lemma \ref{LJWO}:

\begin{proofof}{Lemma \ref{L1inP}}
We construct the randomized unitary $\{q^{(n)}_{j,\delta'},V^{(n)}_{j,\delta'}\}$ satysfying \eqref{awe1'}--\eqref{awe4'} concretely.
Firstly, we label the energy eigenstates of $H^{(n)}_{S|\vec{\lambda}}$ as 
\begin{align}
\ket{\vec{i}_{|\vec{\lambda}}}:=\ket{(i_{1})_{|\lambda_{1}}}\otimes\ket{(i_{2})_{|\lambda_{2}}}\otimes...\otimes\ket{(i_{m})_{|\lambda_{m}}},
\end{align}
where $\ket{(i_{k})_{|\lambda_{k}}}$ is an energy eigenstate of $H^{(n)}_{S_{k}|\lambda_{k}}$ whose engenvalue is the $i_{k}$-th smallest eigenvalue of $H^{(n)}_{k|\lambda_{k}}$.

Secondly, we define the following subspace of $\ca{H}^{(n)}_{S}$ as a function of $\vec{\beta}$, $\vec{\lambda}$ and a real positive number $\epsilon$:$\\$
\textit{Region-$[\vec{\beta},\vec{\lambda},\epsilon]$:}
The subspace of $\ca{H}^{(n)}_{S}$ spanned by $\ket{\vec{i}_{|\vec{\lambda}}}$ whose energy eigenvalues $E^{(n)}_{k|\lambda_{k}}(i_{k})$ of $H^{(n)}_{k|\lambda_{k}}$satisfying
\begin{align}
n\left(\tilde{U}_{k}(\beta_{k},\lambda_{k})-\frac{\epsilon}{m}\right)\le E^{(n)}_{k|\lambda_{k}}(i_{k})\le n\left(\tilde{U}_{k}(\beta_{k},\lambda_{k})+\frac{\epsilon}{m}\right).\Label{A0}
\end{align}
We refer to the projection to the region-$[\vec{\beta},\vec{\lambda},\epsilon]$ and the dimension of the region-$[\vec{\beta},\vec{\lambda},\epsilon]$ as $\Pi^{(n)}_{[\vec{\beta},\vec{\lambda},\epsilon]}$ and $D^{(n)}_{[\vec{\beta},\vec{\lambda},\epsilon]}$.
Then, for a proper real positive number $\alpha$, the projection and dimension  $\Pi^{(n)}_{[\vec{\beta},\vec{\lambda},\epsilon]}$ and $D^{(n)}_{[\vec{\beta},\vec{\lambda},\epsilon]}$ satisfy the following relations for sufficiently large $n$:
\begin{align}
\Tr[\Pi^{(n)}_{[\vec{\beta},\vec{\lambda},\epsilon]}\rho^{(n)}_{\vec{\beta}|\vec{\lambda}}]&\ge1-e^{-n\alpha},\Label{A1}\\
e^{n(\tilde{S}_{S}(\vec{\beta},\vec{\lambda})-2\beta_{\max}\epsilon)}&\le D^{(n)}_{[\vec{\beta},\vec{\lambda},\epsilon]}\le e^{n(\tilde{S}_{S}(\vec{\beta},\vec{\lambda})+2\beta_{\max}\epsilon)}\Label{A2}
\end{align}
(\textit{Proof of \eqref{A1} and \eqref{A2}:}
\eqref{A1} is easily derived from the large deviaiton assumptions \eqref{LDf1} and \eqref{LDf2}. 
Note that the probability $p^{(n)}_{\vec{\beta}|\vec{\lambda}}(\vec{i})$ of $\ket{\vec{i}_{\vec{\lambda}}}$ in $\rho^{(n)}_{\vec{\beta}|\vec{\lambda}}$ satisfies
\begin{align}
p^{(n)}_{\vec{\beta}|\vec{\lambda}}(\vec{i})=e^{-\sum_{k}\beta_{k}(E^{(n)}_{k|\lambda_{k}}(i_{k})-F^{(n)}_{k|\lambda_{k}}(\beta_{k}))},
\end{align}
where $F^{(n)}_{k|\lambda_{k}}(\beta_{k}))$ is the Helmholtz free energy of $\rho^{(n)}_{\beta_{k}|\lambda_{k}}$.
Because of the above and \eqref{A0}, the following inequalities hold:
\begin{align}
e^{-\sum_{k}\beta_{k}(n\tilde{U}_{k}(\beta_{k},\lambda_{k})-F^{(n)}_{k|\lambda_{k}}(\beta_{k}))-\beta_{\max}\epsilon)}\le p^{(n)}_{\vec{\beta}|\vec{\lambda}}(\vec{i})\le e^{-\sum_{k}\beta_{k}(n\tilde{U}_{k}(\beta_{k},\lambda_{k})-F^{(n)}_{k|\lambda_{k}}(\beta_{k}))+\beta_{\max}\epsilon)}.\Label{hasami1}
\end{align}
Because of \eqref{A1}, we obtain 
\begin{align}
1\ge \Tr[\Pi^{(n)}_{[\vec{\beta},\vec{\lambda},\epsilon]}\rho^{(n)}_{\vec{\beta}|\vec{\lambda}}]=\sum_{\vec{i}:\ket{\vec{i}_{|\vec{\lambda}}}\in \mathrm{Region}[\vec{\beta},\vec{\lambda},\epsilon]} p^{(n)}_{\vec{\beta}|\vec{\lambda}}(\vec{i}) \ge1-e^{-n\alpha}\Label{hasami2}
\end{align}
Because of \eqref{hasami1}, \eqref{hasami2} and $\sum_{k}\beta_{k}(n\tilde{U}_{k}(\beta_{k},\lambda_{k})-F^{(n)}_{k|\lambda_{k}}(\beta_{k}))=n\tilde{S}_{S}(\vec{\beta},\vec{\lambda})+o(n)$, the inequalities \eqref{A2} hold for sufficiently large $n$.
)

Now, let us take two real positive numbers $\epsilon$ and $\epsilon'$ satisfying $2\beta''_{\max}(\epsilon+\epsilon')<\delta'$.
Because of \eqref{A2}, $\tilde{S}_{S}(\vec{\beta},\vec{\lambda})<\tilde{S}_{S}(\vec{\beta}',\vec{\lambda}')$ and $\delta'<\tilde{S}_{S}(\vec{\beta}',\vec{\lambda}')-\tilde{S}_{S}(\vec{\beta},\vec{\lambda})$,
the following inequality holds for sufficiently large $n$:
\begin{align}
D^{(n)}_{[\vec{\beta},\vec{\lambda},\epsilon]}<D^{(n)}_{[\vec{\beta}',\vec{\lambda}',\epsilon']}.
\end{align}

Here, we number the energy pure eigenstates $\{\ket{\vec{i}_{|\vec{\lambda}}}\otimes\ket{\vec{\lambda}}\}$ of $H^{(n)}_{I}$ as $\{\ket{x}_{I}\}$ satisfying the following relations:
\begin{align}
\ket{x}_{I}\in (\mathrm{Region}-[\vec{\beta},\vec{\lambda},\epsilon])\otimes\ket{\vec{\lambda}}&\Rightarrow 1\le x\le D^{(n)}_{[\vec{\beta},\vec{\lambda},\epsilon]}\\
\ket{x}_{I}\in (\mathrm{Region}-[\vec{\beta}',\vec{\lambda}',\epsilon'])\otimes\ket{\vec{\lambda}'}&\Rightarrow 1+D^{(n)}_{[\vec{\beta},\vec{\lambda},\epsilon]}\le x\le D^{(n)}_{[\vec{\beta},\vec{\lambda},\epsilon]}+D^{(n)}_{[\vec{\beta}',\vec{\lambda}',\epsilon']}.
\end{align}
(Because $\ket{\lambda}$ and $\ket{\lambda}'$ are orthogonal with each others, this numbering is possible.)

Next, we give the following approximate states of $\rho^{(n)}_{\vec{\beta}|\vec{\lambda}}$ and $\rho^{(n)}_{\vec{\beta}'|\vec{\lambda}'}$:
\begin{align}
\tilde{\rho}^{(n)}_{\vec{\beta}|\vec{\lambda}}&:=\frac{\Pi^{(n)}_{[\vec{\beta},\vec{\lambda},\epsilon]}  \rho^{(n)}_{\vec{\beta}|\vec{\lambda}}  \Pi^{(n)}_{[\vec{\beta},\vec{\lambda},\epsilon]}    }{\Tr[\Pi^{(n)}_{[\vec{\beta},\vec{\lambda},\epsilon]}  \rho^{(n)}_{\vec{\beta}|\vec{\lambda}}  ] }\\
\tilde{\rho}^{(n)}_{\vec{\beta}'|\vec{\lambda}'}&:=\frac{\Pi^{(n)}_{[\vec{\beta}',\vec{\lambda}',\epsilon']}  \rho^{(n)}_{\vec{\beta}'|\vec{\lambda}'} \Pi^{(n)}_{[\vec{\beta}',\vec{\lambda}',\epsilon']}       }{\Tr[\Pi^{(n)}_{[\vec{\beta}',\vec{\lambda}',\epsilon']}  \rho^{(n)}_{\vec{\beta}'|\vec{\lambda}'}  ] }
\end{align}
We also define
\begin{align}
\tilde{\tilde{\rho}}^{(n)}_{\vec{\beta}|\vec{\lambda}}\otimes\ket{\vec{\lambda}'}\bra{\vec{\lambda}'}:=\sum_{x}\ket{f^{(n)}_{0,\delta'}(x)}\bra{x}(\tilde{\rho}^{(n)}_{\vec{\beta}|\vec{\lambda}}\otimes\ket{\vec{\lambda}}\bra{\vec{\lambda}})\ket{x}\bra{f^{(n)}_{0,\delta'}(x)}
\end{align}
where $f^{(n)}_{0,\delta'}(x)$ is an invertible function such that
\begin{align}
f^{(n)}_{0,\delta'}(x)
=\left\{ \begin{array}{ll}
x+D^{(n)}_{[\vec{\beta},\vec{\lambda},\epsilon]} & (1\le x\le D^{(n)}_{[\vec{\beta},\vec{\lambda},\epsilon]}) \\
x-D^{(n)}_{[\vec{\beta},\vec{\lambda},\epsilon]} & (1+D^{(n)}_{[\vec{\beta},\vec{\lambda},\epsilon]}\le x\le 2D^{(n)}_{[\vec{\beta},\vec{\lambda},\epsilon]})\\
x & (2D^{(n)}_{[\vec{\beta},\vec{\lambda},\epsilon]}< x)\\
\end{array} \right..
\end{align}

Note that $\rho^{(n)}_{\vec{\beta}|\vec{\lambda}}$ and $\tilde{\rho}^{(n)}_{\vec{\beta}|\vec{\lambda}}$ have the same entropy spectrum $\tilde{S}_{S}(\vec{\beta},\vec{\lambda})$, and 
$\rho^{(n)}_{\vec{\beta}'|\vec{\lambda}'}$ and $\tilde{\rho}^{(n)}_{\vec{\beta}'|\vec{\lambda}'}$ have the same entropy spectrum $\tilde{S}_{S}(\vec{\beta}',\vec{\lambda}')$.
Also, $\tilde{\tilde{\rho}}^{(n)}_{\vec{\beta}|\vec{\lambda}}$ and $\tilde{\rho}^{(n)}_{\vec{\beta}|\vec{\lambda}}$ has the same entropy spectrum $\tilde{S}_{S}(\vec{\beta},\vec{\lambda})$.
Therefore, the entropy spectrum of $\tilde{\tilde{\rho}}^{(n)}_{\vec{\beta}|\vec{\lambda}}$ is smaller than that of  $\tilde{\rho}^{(n)}_{\vec{\beta}'|\vec{\lambda}'}$.
Also, both of the supports of the states $\tilde{\tilde{\rho}}^{(n)}_{\vec{\beta}|\vec{\lambda}}\otimes\ket{\vec{\lambda}'}\bra{\vec{\lambda}'}$ and $\tilde{\rho}^{(n)}_{\vec{\beta}'|\vec{\lambda}'}\otimes\ket{\vec{\lambda}'}\bra{\vec{\lambda}'}$ are on the region-$[\vec{\beta}',\vec{\lambda}',\epsilon']\otimes\ket{\vec{\lambda}'}$ and dyagonalized the same basis $\{\ket{x}_{I}\}^{D^{(n)}_{[\vec{\beta},\vec{\lambda},\epsilon]}+D^{(n)}_{[\vec{\beta}',\vec{\lambda}',\epsilon']}}_{x=1+D^{(n)}_{[\vec{\beta},\vec{\lambda},\epsilon]}}$  of the region-$[\vec{\beta}',\vec{\lambda}',\epsilon']\otimes\ket{\vec{\lambda}'}$.
Therefore, Lemma \ref{LJWO} guarantees that there exist the probability $\{p^{(n)}_{j,\delta'}\}$ and the invertible functions $\{f^{(n)}_{j,\delta'}\}$ from $\{x\}^{D^{(n)}_{[\vec{\beta},\vec{\lambda},\epsilon]}+D^{(n)}_{[\vec{\beta}',\vec{\lambda}',\epsilon']}}_{x=1+D^{(n)}_{[\vec{\beta},\vec{\lambda},\epsilon]}}$ to $\{x\}^{D^{(n)}_{[\vec{\beta},\vec{\lambda},\epsilon]}+D^{(n)}_{[\vec{\beta}',\vec{\lambda}',\epsilon']}}_{x=1+D^{(n)}_{[\vec{\beta},\vec{\lambda},\epsilon]}}$ such that
\begin{align}
\lim_{n\rightarrow\infty}\|\tilde{\tilde{\rho}}^{(n)}_{\vec{\beta}'|\vec{\lambda}'}\otimes\ket{\vec{\lambda}'}\bra{\vec{\lambda}'}
-\tilde{\rho}^{(n)}_{\vec{\beta}'|\vec{\lambda}'}\otimes\ket{\vec{\lambda}'}\bra{\vec{\lambda}'}\|&=0,\Label{A3}\\
\mathrm{supp}[\tilde{\tilde{\rho}}^{(n)}_{\vec{\beta}'|\vec{\lambda}'}]&\subset \mathrm{region}-[\vec{\beta}',\vec{\lambda}',\epsilon'],
\end{align}
where
\begin{align}
\tilde{\tilde{\rho}}^{(n)}_{\vec{\beta}'|\vec{\lambda}'}\otimes\ket{\vec{\lambda}'}\bra{\vec{\lambda}'} =\sum_{j}p^{(n)}_{j,\delta'}\sum_{x}\ket{f^{(n)}_{j,\delta'}(x)}\bra{x}
(\tilde{\tilde{\rho}}^{(n)}_{\vec{\beta}|\vec{\lambda}}\otimes\ket{\vec{\lambda}'}\bra{\vec{\lambda}'})
\ket{x}\bra{f^{(n)}_{j,\delta'}(x)}.
\end{align}

Now, we give $\{q^{(n)}_{j,\delta'},V^{(n)}_{j,\delta'}\}$ that we seek.
They are defined as follows:
\begin{align}
q^{(n)}_{j,\delta'}&:=p^{(n)}_{j,\delta'}\\
V^{(n)}_{j,\delta'}&:=\sum_{x,e}\ket{g^{(n)}_{j,\delta'}(x,e)}_{IE_{W}}\bra{x,e}_{IE_{W}}
\end{align}
where $\ket{x,e}_{IE_{W}}:=\ket{x}_{I}\otimes\ket{e}_{E_{W}}$ and where $g^{(n)}_{j,\delta'}(x,e)$ is an invertible function defined as
\begin{align}
g^{(n)}_{j,\delta'}(x,e):=
\left\{ \begin{array}{ll}
\left(f^{(n)}_{j,\delta'}\circ f^{(n)}_{0,\delta'}(x), e+E^{(n)}_{x}-E^{(n)}_{f^{(n)}_{j,\delta'}\circ f^{(n)}_{0,\delta'}(x)}\right) & (n(\tilde{U}_{W}-|\Delta\tilde{U}_{S}|-4h_{0})\le e\le n(\tilde{U}_{W}+|\Delta\tilde{U}_{S}|+4h_{0})) \\
(x,e) & (\mathrm{Otherwise})\\
\end{array} \right..
\end{align}
By definition, each $V^{(n)}_{j,\delta'}$ is energy-preserving.
Therefore, we only have to show that the random unitary satisfies \eqref{awe1'}--\eqref{awe4'}.

We firstly show that $\{q^{(n)}_{j,\delta'},V^{(n)}_{j,\delta'}\}$ satisfies \eqref{awe1'}.
By definition of $g^{(n)}_{j,\delta'}(x,e)$, for arbitrary $e$ satisfying $n(\tilde{U}_{W}-\delta)\le e\le n(\tilde{U}_{W}+\delta)$, the following relation holds:
\begin{align}
\tilde{\tilde{\rho}}^{(n)}_{\vec{\beta}'|\vec{\lambda}'}\otimes\ket{\vec{\lambda}'}\bra{\vec{\lambda}'} =\Tr_{E_{W}}[\sum_{j}q^{(n)}_{j,\delta'}\sum_{x}V^{(n)}_{j,\delta'}
(\tilde{\rho}^{(n)}_{\vec{\beta}|\vec{\lambda}}\otimes\ket{\vec{\lambda}}\bra{\vec{\lambda}}\otimes \ket{e}\bra{e}_{E_{W}})
V^{(n)\dagger}_{j,\delta'}].
\end{align}
Therefore, we obtain
\begin{align}
\tilde{\tilde{\rho}}^{(n)}_{\vec{\beta}'|\vec{\lambda}'}\otimes\ket{\vec{\lambda}'}\bra{\vec{\lambda}'} =\Tr_{E_{W}}[\sum_{j}q^{(n)}_{j,\delta'}\sum_{x}V^{(n)}_{j,\delta'}
(\tilde{\rho}^{(n)}_{\vec{\beta}|\vec{\lambda}}\otimes\ket{\vec{\lambda}}\bra{\vec{\lambda}}\otimes \pi^{(n)}_{\tilde{U}_{W},\delta})\Label{B23}
V^{(n)\dagger}_{j,\delta'}].
\end{align}
Because of \eqref{A1} and the gentle measurement lemma \cite{gentle}, the following inequalities hold for the sufficiently large $n$:
\begin{align}
\|\rho^{(n)}_{\vec{\beta}|\vec{\lambda}}\otimes\ket{\vec{\lambda}}\bra{\vec{\lambda}}-\tilde{\rho}^{(n)}_{\vec{\beta}|\vec{\lambda}}\otimes\ket{\vec{\lambda}}\bra{\vec{\lambda}}\|\le e^{-n\alpha/2}\Label{app1}\\
\|\rho^{(n)}_{\vec{\beta}'|\vec{\lambda}'}\otimes\ket{\vec{\lambda}'}\bra{\vec{\lambda}'}-\tilde{\rho}^{(n)}_{\vec{\beta}'|\vec{\lambda}'}\otimes\ket{\vec{\lambda}'}\bra{\vec{\lambda}'}\|\le e^{-n\alpha/2}\Label{app1'}
\end{align}
From \eqref{A3}, \eqref{B23}, \eqref{app1} and \eqref{app1'}, we obtain
\begin{align}
\lim_{n\rightarrow\infty}\left\|\Tr_{E_{W}}[\sum_{j}q^{(n)}_{j,\delta'}\sum_{x}V^{(n)}_{j,\delta'}
(\rho^{(n)}_{\vec{\beta}|\vec{\lambda}}\otimes\ket{\vec{\lambda}}\bra{\vec{\lambda}}\otimes \pi^{(n)}_{\tilde{U}_{W},\delta})
V^{(n)\dagger}_{j,\delta'}]-\rho^{(n)}_{\vec{\beta}'|\vec{\lambda}'}\otimes\ket{\vec{\lambda}'}\bra{\vec{\lambda}'}\right\|=0.
\end{align}
Because of $\sigma^{(n)}_{I,\delta'}=\Tr_{E_{W}}[\sum_{j}q^{(n)}_{j,\delta'}\sum_{x}V^{(n)}_{j,\delta'}
(\rho^{(n)}_{\vec{\beta}|\vec{\lambda}}\otimes\ket{\vec{\lambda}}\bra{\vec{\lambda}}\otimes \pi^{(n)}_{\tilde{U}_{W},\delta})
V^{(n)\dagger}_{j,\delta'}]$, \eqref{awe1'} holds.

Next, we show that $\{q^{(n)}_{j,\delta'},V^{(n)}_{j,\delta'}\}$ satisfies \eqref{awe3'} and \eqref{awe4'}.
We firstly note that $\sigma^{e,(n)}_{IE_{W}}:=\sum_{j}q^{(n)}_{j,\delta'}\sum_{x}V^{(n)}_{j,\delta'}
(\rho^{(n)}_{\vec{\beta}|\vec{\lambda}}\otimes\ket{\vec{\lambda}}\bra{\vec{\lambda}}\otimes \ket{e}\bra{e}_{E_{W}})
V^{(n)\dagger}_{j,\delta'}$ can be described as follows:
\begin{align}
\sigma^{e,(n)}_{IE_{W}}=\sum_{j}\sum_{x}q^{(n)}_{j,\delta'}p^{(n)}_{\mathrm{ini}}(x)\ketbra{f^{(n)}_{j,\delta'}\circ f^{(n)}_{0,\delta'}(x)}\otimes\ketbra{e+E^{(n)}_{x}-E^{(n)}_{f^{(n)}_{j,\delta'}\circ f^{(n)}_{0,\delta'}(x)}}
\end{align}
Therefore, the reduced state $\sigma^{e,(n)}_{E_{W}}:=\Tr_{I}[\sigma^{e,(n)}_{IE_{W}}]$ is diagonalized with the energy eigenstates of $H^{(n)}_{E_{W}}$
Thus, with a proper probability distribution $\{r^{(n)}_{e'}\}$,  the reduced state $\sigma^{e,(n)}_{E_{W}}:=\Tr_{I}[\sigma^{e,(n)}_{IE_{W}}]$ can be described as follows:
\begin{align}
\sigma^{e,(n)}_{E_{W}}=\sum_{e'}r^{(n)}_{e'}\ket{e-n\Delta\tilde{U}_{S}+e'}\bra{e-n\Delta\tilde{U}_{S}+e'}.
\end{align}
As shown below, the distribution $\{r^{(n)}_{e'}\}$ satisfies
\begin{align}
r^{(n)}_{\mathrm{main}}&=\Tr[\Pi^{(n)}_{[\vec{\beta},\vec{\lambda},\epsilon]}\rho^{(n)}_{\vec{\beta}|\vec{\lambda}}]\ge1-e^{-n\alpha}\Label{rcond}\\
(1-r^{(n)}_{\mathrm{main}})H\{\tilde{r}^{(n)}_{e'}\}&=-(1-r^{(n)}_{\mathrm{main}})\Tr[\rho^{(n)\lnot}_{\vec{\beta}|\vec{\lambda}}\log\rho^{(n)\lnot}_{\vec{\beta}|\vec{\lambda}}]=o(n)\Label{rcond2}.
\end{align}
where
\begin{align}
r^{(n)}_{\mathrm{main}}&:=\sum_{-n(\epsilon+\epsilon')\le e'\le n(\epsilon+\epsilon')}r^{(n)}_{e'},\\
\tilde{r}^{(n)}_{e'}&:=\frac{r^{(n)}_{e'}}{1-r^{(n)}_{\mathrm{main}}},\enskip \mathrm{for}\enskip e'\notin(-n(\epsilon+\epsilon'), n(\epsilon+\epsilon'))\\
\rho^{(n)\lnot}_{\vec{\beta}|\vec{\lambda}}&:=\frac{(1-\Pi^{(n)}_{[\vec{\beta},\vec{\lambda},\epsilon]})\rho^{(n)}_{\vec{\beta}|\vec{\lambda}}(1-\Pi^{(n)}_{[\vec{\beta},\vec{\lambda},\epsilon]})}{\Tr[(1-\Pi^{(n)}_{[\vec{\beta},\vec{\lambda},\epsilon]})\rho^{(n)}_{\vec{\beta}|\vec{\lambda}}]}.
\end{align}
(\textit{Proof of \eqref{rcond} and \eqref{rcond2}:}
\eqref{rcond} is easily derived from the large deviation assumptions \eqref{LDf1} and \eqref{LDf2}.
\eqref{rcond2} is given in the same way as the derivation of \eqref{entropysub}:
\begin{align}
S(\rho^{(n)}_{\vec{\beta}|\vec{\lambda}})&=S(r^{(n)}_{\mathrm{main}}\tilde{\rho}^{(n)}_{\vec{\beta}|\vec{\lambda}}+(1-r^{(n)}_{\mathrm{main}})\rho^{(n)\lnot}_{\vec{\beta}|\vec{\lambda}})\ge r^{(n)}_{\mathrm{main}}S(\tilde{\rho}^{(n)}_{\vec{\beta}|\vec{\lambda}})+(1-r^{(n)}_{\mathrm{main}})S(\rho^{(n)\lnot}_{\vec{\beta}|\vec{\lambda}})\nonumber\\
&\ge r^{(n)}_{\mathrm{main}}(n\tilde{S}(\vec{\beta},\vec{\lambda})+o(n)-e^{-n\alpha/2}\log e^{n(\tilde{S}_{S}(\vec{\beta},\vec{\lambda})+2\beta''_{\mathrm{max}}\epsilon)})+(1-r^{(n)}_{\mathrm{main}})S(\rho^{(n)\lnot}_{\vec{\beta}|\vec{\lambda}}).
\end{align}

Because $\pi^{(n)}_{\tilde{U}_{W},\delta}$ is described as 
\begin{align}
\pi^{(n)}_{\tilde{U}_{W},\delta}=\frac{1}{C}\sum_{-n(\tilde{U}_{W}-\delta)\le e\le n(\tilde{U}_{W}+\delta)}\ket{e}\bra{e},
\end{align}
the state $\sigma^{(n)}_{E_{W}}:=\Tr_{I}[\sigma^{(n)}_{IE_{W}}]$ is described as follows:
\begin{align}
\sigma^{(n)}_{E_{W}}&=\sum_{n(\tilde{U}_{W}-\delta)\le e\le n(\tilde{U}_{W}+\delta)}\frac{1}{C}\sum_{e'}r^{(n)}_{e'}\ket{e-n\Delta\tilde{U}_{S}+e'}\bra{e-n\Delta\tilde{U}_{S}+e'}\\
&=\sum_{e'}r^{(n)}_{e'}\sum_{n(\tilde{U}_{W}-\delta)\le e\le n(\tilde{U}_{W}+\delta)}\frac{1}{C}\ket{e-n\Delta\tilde{U}_{S}+e'}\bra{e-n\Delta\tilde{U}_{S}+e'}\\
&=\sum_{e'}r^{(n)}_{e'}\pi^{(n)}_{\tilde{U}_{W}-\Delta\tilde{U}_{S}+\frac{e'}{n},\delta}
\end{align}
Therefore, 
\begin{align}
S(\sigma^{(n)}_{E_{W}})\ge\sum_{e'}r^{(n)}_{e'}S(\pi^{(n)}_{\tilde{U}_{W}-\Delta\tilde{U}_{S}+\frac{e'}{n},\delta})=S(\pi^{(n)}_{\tilde{U}_{W},\delta}).
\end{align}
Thus, in order to show \eqref{awe4'}, we only have to show $S(\sigma^{(n)}_{E_{W}})\le S(\pi^{(n)}_{\tilde{U}_{W},\delta})+o(n).$

Let us show $S(\sigma^{(n)}_{E_{W}})\le S(\pi^{(n)}_{\tilde{U}_{W},\delta})+o(n).$
With using
\begin{align}
\pi^{(n)}_{\mbox{main}}:=\frac{\sum_{-n(\epsilon+\epsilon')\le e'\le n(\epsilon+\epsilon')}r^{(n)}_{e'}\pi^{(n)}_{\tilde{U}_{W}-\Delta\tilde{U}_{S}+\frac{e'}{n},\delta}}{r^{(n)}_{\mathrm{main}}},
\end{align}
the state $\sigma^{(n)}_{E_{W}}$ is described as
\begin{align}
\sigma^{(n)}_{E_{W}}=r^{(n)}_{\mathrm{main}}\pi^{(n)}_{\mathrm{main}}+(1-r^{(n)}_{\mathrm{main}})\sum_{e'\notin(-n(\epsilon+\epsilon'), n(\epsilon+\epsilon'))}\tilde{r}^{(n)}_{e'}\pi^{(n)}_{\tilde{U}_{W}-\Delta\tilde{U}_{S}+\frac{e'}{n},\delta}\Label{decom1}.
\end{align}
Therefore, $S(\sigma^{(n)}_{E_{W}})$ satisfies
\begin{align}
S(\sigma^{(n)}_{E_{W}})&\le H\{r^{(n)}_{\mathrm{main}},1-r^{(n)}_{\mathrm{main}}\}+r^{(n)}_{\mathrm{main}}S(\pi^{(n)}_{\mathrm{main}})+(1-r^{(n)}_{\mathrm{main}})S(\sum_{e'\notin(-n(\epsilon+\epsilon'), n(\epsilon+\epsilon'))}\tilde{r}^{(n)}_{e'}\pi^{(n)}_{\tilde{U}_{W}-\Delta\tilde{U}_{S}+\frac{e'}{n},\delta})\nonumber\\
&\le H\{r^{(n)}_{\mathrm{main}},1-r^{(n)}_{\mathrm{main}}\}+r^{(n)}_{\mathrm{main}}S(\pi^{(n)}_{\mathrm{main}})+(1-r^{(n)}_{\mathrm{main}})\left(H\{\tilde{r}^{(n)}_{e'}\}+\sum_{e'\notin(-n(\epsilon+\epsilon'), n(\epsilon+\epsilon'))}\tilde{r}^{(n)}_{e'}S(\pi^{(n)}_{\tilde{U}_{W}-\Delta\tilde{U}_{S}+\frac{e'}{n},\delta})\right)\nonumber\\
&= H\{r^{(n)}_{\mathrm{main}},1-r^{(n)}_{\mathrm{main}}\}+r^{(n)}_{\mathrm{main}}S(\pi^{(n)}_{\mathrm{main}})+(1-r^{(n)}_{\mathrm{main}})\left(H\{\tilde{r}^{(n)}_{e'}\}+\sum_{e'\notin(-n(\epsilon+\epsilon'), n(\epsilon+\epsilon'))}\tilde{r}^{(n)}_{e'}S(\pi^{(n)}_{\tilde{U}_{W}-\Delta\tilde{U}_{S},\delta})\right)\nonumber\\
&\stackrel{(a)}{\le} \log 2+r^{(n)}_{\mathrm{main}}S(\pi^{(n)}_{\mathrm{main}})+o(n)+(1-r^{(n)}_{\mathrm{main}})\sum_{e'\notin(-n(\epsilon+\epsilon'), n(\epsilon+\epsilon'))}\tilde{r}^{(n)}_{e'}S(\pi^{(n)}_{\tilde{U}_{W}-\Delta\tilde{U}_{S},\delta})\nonumber\\
&= \log 2+r^{(n)}_{\mathrm{main}}S(\pi^{(n)}_{\mathrm{main}})+(1-r^{(n)}_{\mathrm{main}})S(\pi^{(n)}_{\tilde{U}_{W},\delta})+o(n),
\end{align}
where $(a)$ is given by $H\{r^{(n)}_{\mathrm{main}},1-r^{(n)}_{\mathrm{main}}\}\le\log2$ and \eqref{rcond2}.
Let us decompose $\pi^{(n)}_{\mathrm{main}}$ as follows:
\begin{align}
\pi^{(n)}_{\mathrm{main}}=r^{(n)}_{A}\pi^{(n)}_{A}+r^{(n)}_{B}\pi^{(n)}_{B}
\end{align}
where
\begin{align}
r^{(n)}_{A}&:=\Tr[\Pi^{(n)}_{A}\pi^{(n)}_{\mathrm{main}}]\\
\pi^{(n)}_{A}&:=\frac{\Pi^{(n)}_{A}\pi^{(n)}_{\mathrm{main}}\Pi^{(n)}_{A}}{r^{(n)}_{A}}\\
\Pi^{(n)}_{A}&:=\sum_{n(\tilde{U}_{W}-\Delta\tilde{U}_{S}-\delta+\epsilon+\epsilon')\le e\le n(\tilde{U}_{W}-\Delta\tilde{U}_{S}+\delta-\epsilon-\epsilon')}\ket{e}\bra{e}\\
r^{(n)}_{B}&:=\Tr[\Pi^{(n)}_{B}\pi^{(n)}_{\mathrm{main}}]\\
\pi^{(n)}_{B}&:=\frac{\Pi^{(n)}_{B}\pi^{(n)}_{\mathrm{main}}\Pi^{(n)}_{B}}{r^{(n)}_{B}}\\
\Pi^{(n)}_{B}&:=\sum_{n(\tilde{U}_{W}-\Delta\tilde{U}_{S}-\delta-\epsilon-\epsilon')\le e\le n(\tilde{U}_{W}-\Delta\tilde{U}_{S}+\delta+\epsilon+\epsilon')}\ket{e}\bra{e}-\Pi^{(n)}_{A}
\end{align}
Because the support of $\pi^{(n)}_{\mathrm{main}}$ is on the subspace projected by the projection $\Pi^{(n)}_{A}+\Pi^{(n)}_{B}$,  the equality $r^{(n)}_{A}+r^{(n)}_{B}=1$ holds.
Because an arbitrary quantum state $\rho$ has entropy lower than the maximally mixed state on the support of $\rho$, we obtain
\begin{align}
S(\pi^{(n)}_{A})\le S(\pi^{(n)}_{\tilde{U}_{W}-\Delta \tilde{U}_{S},\delta-\epsilon-\epsilon'})\\
S(\pi^{(n)}_{B})\le S(\frac{\pi^{(n)}_{\tilde{U}_{W}-\Delta \tilde{U}_{S}-\delta,\epsilon+\epsilon'}+\pi^{(n)}_{\tilde{U}_{W}-\Delta \tilde{U}_{S}+\delta,\epsilon+\epsilon'}}{2}).
\end{align}
Similarly, because the support of $\pi^{(n)}_{\tilde{U}_{W}-\Delta \tilde{U}_{S},\delta-\epsilon-\epsilon'}$ is included in the support of $\pi^{(n)}_{\tilde{U}_{W}-\Delta \tilde{U}_{S},\delta}$, we obtain
\begin{align}
S(\pi^{(n)}_{\tilde{U}_{W}-\Delta \tilde{U}_{S},\delta-\epsilon-\epsilon'})\le S(\pi^{(n)}_{\tilde{U}_{W}-\Delta \tilde{U}_{S},\delta})=S(\pi^{(n)}_{\tilde{U}_{W},\delta}).
\end{align}
Because of $2\beta''_{\mathrm{max}}(\epsilon+\epsilon')<\delta'<2\beta''_{\mathrm{max}}\delta$, we obtain
\begin{align}
S(\frac{\pi^{(n)}_{\tilde{U}_{W}-\Delta \tilde{U}_{S}-\delta,\epsilon+\epsilon'}+\pi^{(n)}_{\tilde{U}_{W}-\Delta \tilde{U}_{S}+\delta,\epsilon+\epsilon'}}{2})
&\le H\{\frac{1}{2},\frac{1}{2}\}+\frac{1}{2}S(\pi^{(n)}_{\tilde{U}_{W}-\Delta \tilde{U}_{S}-\delta,\epsilon+\epsilon'})+\frac{1}{2}S(\pi^{(n)}_{\tilde{U}_{W}-\Delta \tilde{U}_{S}+\delta,\epsilon+\epsilon'})\\
&\le H\{\frac{1}{2},\frac{1}{2}\}+\frac{1}{2}S(\pi^{(n)}_{\tilde{U}_{W}-\Delta \tilde{U}_{S}-\delta,\delta})+\frac{1}{2}S(\pi^{(n)}_{\tilde{U}_{W}-\Delta \tilde{U}_{S}+\delta,\delta})\nonumber\\
&=\log 2+S(\pi^{(n)}_{\tilde{U}_{W},\delta}).
\end{align}
Therefore, we obtain
\begin{align}
S(\pi^{(n)}_{\mathrm{main}})=H\{r^{(n)}_{A},r^{(n)}_{B}\}+r^{(n)}_{A}S(\pi^{(n)}_{A})+r^{(n)}_{B}S(\pi^{(n)}_{B})\le2\log2+S(\pi^{(n)}_{\tilde{U}_{W},\delta})
\end{align}
Therefore,
\begin{align}
S(\sigma^{(n)}_{E_{W}})&\le 3\log 2+S(\pi^{(n)}_{\tilde{U}_{W},\delta})+o(n)
\end{align}
holds.
Thus, $\{q^{(n)}_{j,\delta'},V^{(n)}_{j,\delta'}\}$ satisfies \eqref{awe4'}.

Finally, we show that $\{q^{(n)}_{j,\delta'},V^{(n)}_{j,\delta'}\}$ satisfies \eqref{awe3'}.
Note that $\pi^{(n)}_{A}$ is the maximally mixed state on the subspace projected by $\Pi^{(n)}_{A}$ and that $\Tr[\Pi^{(n)}_{A}\pi^{(n)}_{\tilde{U}_{W}-\Delta \tilde{U}_{S},\delta}]=\frac{\delta-(\epsilon+\epsilon')}{\delta}$.
Therefore, $\pi^{(n)}_{\tilde{U}_{W}-\Delta \tilde{U}_{S},\delta}$ is described as
\begin{align}
\pi^{(n)}_{\tilde{U}_{W}-\Delta\tilde{U}_{S},\delta}&=\frac{\delta-(\epsilon+\epsilon')}{\delta}\pi^{(n)}_{A}+(1-\frac{\delta-(\epsilon+\epsilon')}{\delta})\pi^{(n)}_{\tilde{U}_{W}-\Delta\tilde{U}_{S},\delta,B}\nonumber\\
&=\frac{\delta-(\epsilon+\epsilon')}{\delta}\pi^{(n)}_{A}+\frac{\epsilon+\epsilon'}{\delta}\pi^{(n)}_{\tilde{U}_{W}-\Delta\tilde{U}_{S},\delta,B},
\end{align}
where
\begin{align}
\pi^{(n)}_{\tilde{U}_{W}-\Delta\tilde{U}_{S},\delta,B}:=\frac{\Pi^{(n)}_{B}\pi^{(n)}_{\tilde{U}_{W}-\Delta\tilde{U}_{S},\delta}\Pi^{(n)}_{B}}{\Tr[\Pi^{(n)}_{B}\pi^{(n)}_{\tilde{U}_{W}-\Delta\tilde{U}_{S},\delta}]}.
\end{align}
Therefore, we obtain
\begin{align}
\|\pi^{(n)}_{\mathrm{main}}-\pi^{(n)}_{\tilde{U}_{W}-\Delta\tilde{U}_{S},\delta}\|
\le 2\left|r^{(n)}_{A}-\frac{\delta-(\epsilon+\epsilon')}{\delta}\right|+\left\|r^{(n)}_{B}\pi^{(n)}_{B}-\frac{\epsilon+\epsilon'}{\delta}\pi^{(n)}_{\tilde{U}_{W}-\Delta\tilde{U}_{S},\delta}\right\|
\le 2\left|r^{(n)}_{A}-\frac{\delta-(\epsilon+\epsilon')}{\delta}\right|+2(r^{(n)}_{B}+\frac{\epsilon+\epsilon'}{\delta}).
\end{align}
Because $\pi^{(n)}_{\mathrm{main}}$ and $\pi^{(n)}_{\tilde{U}_{W}-\Delta\tilde{U}_{S},\delta}$ have the same support,
and because $\pi^{(n)}_{\tilde{U}_{W}-\Delta\tilde{U}_{S},\delta}$ is the maximally mixed state on the support,  the inequality
\begin{align}
(1-r^{(n)}_{A})=\Tr[(1-\Pi^{(n)}_{A})\pi^{(n)}_{\mathrm{main}}]\le\Tr[(1-\Pi^{(n)}_{A})\pi^{(n)}_{\tilde{U}_{W}-\Delta\tilde{U}_{S},\delta}]
=(1-\frac{\delta-\epsilon-\epsilon'}{\delta})
\end{align}
holds.
Because of $r^{(n)}_{A}+r^{(n)}_{B}=1$, the inequality $r^{(n)}_{B}\le\frac{\epsilon+\epsilon'}{\delta}\le\frac{\delta'}{2\beta''_{\mathrm{max}}\delta}$ holds.
Therefore, 
\begin{align}
\|\pi^{(n)}_{\mathrm{main}}-\pi^{(n)}_{\tilde{U}_{W}-\Delta\tilde{U}_{S},\delta}\|
&\le 2\left|r^{(n)}_{A}-\frac{\delta-(\epsilon+\epsilon')}{\delta}\right|+2(r^{(n)}_{B}+\frac{\epsilon+\epsilon'}{\delta})\nonumber\\
&\le 2\left|1-\frac{\delta-(\epsilon+\epsilon')}{\delta}\right|+2(r^{(n)}_{B}+\frac{\epsilon+\epsilon'}{\delta})\nonumber\\
&\le 2\left|1-\frac{\delta-(\epsilon+\epsilon')}{\delta}\right|+\frac{2\delta'}{\beta''_{\mathrm{max}}\delta}\nonumber\\
&\le \frac{4\delta'}{\beta''_{\max}\delta}
\end{align}
Because of \eqref{rcond} and \eqref{decom1}, we obtain
\begin{align}
\|\sigma^{(n)}_{E_{W}}-\pi^{(n)}_{\mathrm{main}}\|\le O(e^{-n\alpha}).
\end{align}
Hence, 
\begin{align}
\lim_{n\rightarrow\infty}\|\sigma^{(n)}_{E_{W}}-\pi^{(n)}_{\tilde{U}_{W}-\Delta\tilde{U}_{S},\delta}\|\le\frac{4\delta'}{\beta''_{\mathrm{max}}\delta}
\end{align}
Therefore, $\{q^{(n)}_{j,\delta'},V^{(n)}_{j,\delta'}\}$ satisfies \eqref{awe3'}.

\end{proofof}

\section{Proof of Corollaries \ref{PM} and \ref{1st}}\Label{PC1}

\begin{proof}
Let us take $\beta''_{M}$ and $\tilde{W}''_{M}$ arbitrarily.
Then, the adiabatic work extraction $(\beta,\lambda_{S})\times(\beta,\lambda_{B})\rightarrow_{\mathrm{ad}:\tilde{W}''_{M}}(\beta,\lambda'_{S})\times(\beta''_{M},\lambda_{B})$ is possible if and only if the following two expressions hold:
\begin{align}
\tilde{S}_{S}(\beta,\lambda_{S})+\tilde{S}_{B_{M}}(\beta,\lambda_{B})&\le\tilde{S}_{S}(\beta,\lambda'_{S})+\tilde{S}_{B_{M}}(\beta''_{M},\lambda_{B}),\Label{C11}\\
\tilde{W}''_{M}&=-\Delta \tilde{U}_{S}+\tilde{Q}''_{M},\Label{C12}
\end{align}
where $\tilde{Q}''_{M}:=\tilde{U}_{B_{M}}(\beta,\lambda_{B})-\tilde{U}_{B_{M}}(\beta''_{M},\lambda_{B})$.
Substituting \eqref{SUF} into \eqref{C11} and \eqref{C12}, we obtain the following neceesary and sufficient condition of the possibility of the adiabatic work extraction $(\beta,\lambda_{S})\times(\beta,\lambda_{B})\rightarrow_{\mathrm{ad}:\tilde{W}''_{M}}(\beta,\lambda'_{S})\times(\beta''_{M},\lambda_{B})$:
\begin{align}
\tilde{W}''_{M}&\le-\Delta \tilde{F}_{S}-\frac{\int^{\tilde{U}_{B_{M}}(\beta,\lambda_{B})-X_{M}}_{\tilde{U}_{B_{M}}(\beta,\lambda_{B})}\beta_{M}(u)du-\beta\int^{\tilde{U}_{B_{M}}(\beta,\lambda_{B})-X_{M}}_{\tilde{U}_{B_{M}}(\beta,\lambda_{B})}du}{\beta}\Label{C11'}
\\
\tilde{W}''_{M}&=-\Delta \tilde{U}_{S}+\tilde{Q}''_{M}\Label{C12'},
\end{align}
where $\beta_{M}(u)$ is the reverse function of $\tilde{U}_{B_{M}}(\beta,\lambda_{B})$ (Note that $\lambda_{B}$ is fixed.), and where
\begin{align}
X_{M}&:=\tilde{W}''_{M}+\Delta \tilde{U}_{S}.
\end{align}
Note that $\tilde{X}$ is not equal to $\tilde{Q}''_{M}$ in general, although they are equal to each others when \eqref{C12'} holds.
Moreover, the second term of the righthand side of \eqref{C11'} satisfies
\begin{align}
\left|\frac{\int^{\tilde{U}_{B_{M}}(\beta,\lambda_{B})-X_{M}}_{\tilde{U}_{B_{M}}(\beta,\lambda_{B})}\beta_{M}(u)du-\beta\int^{\tilde{U}_{B_{M}}(\beta,\lambda_{B})-X_{M}}_{\tilde{U}_{B_{M}}(\beta,\lambda_{B})}du}{\beta}\right|\le\frac{|\beta-\beta''_{M}|}{\beta}|X_{M}|.\Label{CCC}
\end{align}

Now we have completed the preparation.
We firstly prove Corollary \ref{PM}.
When $\tilde{W}\le-\Delta \tilde{F}_{S}$ holds, we define $\tilde{W}_{M}$ and $\tilde{\beta}'_{M}$ as follows:
\begin{align}
\tilde{W}_{M}&:=\tilde{W}-\frac{1}{\sqrt{M}},\\
\tilde{\beta}'_{M}&:=\beta_{M}(\tilde{U}_{B_{M}}(\beta,\lambda_{M})-\tilde{W}_{M}-\Delta \tilde{U}_{S})
\end{align}
Clearly, $\tilde{W}_{M}$ satisfies $\tilde{W}_{M}=\tilde{W}+O(1/\sqrt{M})$.
Because the specific heat $B_{M}$ is proportional to $M$, $\tilde{\beta}'_{M}=\beta+O(1/M)$ holds.
Therefore, because of \eqref{CCC}, the order of the second term of the righthand side of \eqref{C12'} is $O(1/M)$.
Hence, when we define the functions $\tilde{W}_{M}$ and $\beta'_{M}$ as $\tilde{W}''_{M}$ and $\beta''_{M}$,
$\tilde{W}_{M}$ and $\beta'_{M}$ satisfy \eqref{C11'} and \eqref{C12'} for sufficiently large $M$.
Therefore, there exist $\tilde{W}_{M}$ and $\beta'_{M}$ satisfying $\lim_{M\rightarrow\infty}\tilde{W}_{M}=\tilde{W}$ and $\lim_{M\rightarrow\infty}\beta'_{M}=\beta$, and 
the adiabatic work extraction $(\beta,\lambda_{S})\times(\beta,\lambda_{B})\rightarrow_{\mathrm{ad}:\tilde{W}_{M}}(\beta,\lambda'_{S})\times(\beta'_{M},\lambda_{B})$
is possible for sufficiently large $M$.
Therefore, when $\tilde{W}\le-\Delta \tilde{F}_{S}$ holds, 
the isothermal work extraction $(\beta,\lambda_{S})\rightarrow_{\mathrm{is}:\tilde{W}}(\beta,\lambda'_{S})$ is possible.

Next, let us show that when $\tilde{W}>-\Delta \tilde{F}_{S}$ holds, the isothermal work extraction $(\beta,\lambda_{S})\rightarrow_{\mathrm{is}:\tilde{W}}(\beta,\lambda'_{S})$ is impossible.
We define a number $\delta>0$ as $\delta:=\tilde{W}+\Delta \tilde{F}_{S}$.
We also take arbitrary functions $\tilde{W}_{M}$ and $\beta'_{M}$ satisfying $\tilde{W}_{M}=\tilde{W}+o(1)$ and $\beta'_{M}=\beta+o(1)$.
Substituting $\tilde{W}''_{M}$ and $\beta''_{M}$ for \eqref{C11'} and \eqref{C12'}, we obtain that the order of the second term of the righthand side of \eqref{C12'} is $o(1)$.
On the other hand, the lefthand side of \eqref{C12'} is equal to $-\Delta F_{S}+\delta+o(1)$.
Therefore, \eqref{C12'} is invalid for sufficiently large $M$.
Therefore, no matter what $\tilde{W}_{M}$ and $\beta'_{M}$ satisfying $\tilde{W}_{M}=\tilde{W}+o(1)$ and $\beta'_{M}=\beta+o(1)$ are used, 
the adiabatic work extraction $(\beta,\lambda_{S})\times(\beta,\lambda_{B})\rightarrow_{\mathrm{ad}:\tilde{W}_{M}}(\beta,\lambda'_{S})\times(\beta'_{M},\lambda_{B})$ is impossible for sufficiently large $M$.
Therefore, when $\tilde{W}>-\Delta \tilde{F}_{S}$ holds, the isothermal work extraction $(\beta,\lambda_{S})\rightarrow_{\mathrm{is}:\tilde{W}}(\beta,\lambda'_{S})$ is impossible.

Finally, let us show Corollary \ref{1st}.
When $(\beta,\lambda_{S})\rightarrow_{\mathrm{is}:\tilde{W}}(\beta,\lambda'_{S})$ is possible,
there exist two functions $\tilde{W}_{M}$ and $\beta'_{M}$ satisfying $\tilde{W}_{M}=\tilde{W}+o(1)$ and $\beta'_{M}=\beta+o(1)$,
and the adiabatic work extraction $(\beta,\lambda_{S})\times(\beta,\lambda_{B})\rightarrow_{\mathrm{ad}:\tilde{W}_{M}}(\beta,\lambda'_{S})\times(\beta'_{M},\lambda_{B})$ is possible.
Let us take arbitrary set of $\tilde{W}_{M}$ and $\beta'_{M}$ satisfying $\tilde{W}_{M}=\tilde{W}+o(1)$ and $\beta'_{M}=\beta+o(1)$ and satisfying that the adiabatic work extraction $(\beta,\lambda_{S})\times(\beta,\lambda_{B})\rightarrow_{\mathrm{ad}:\tilde{W}_{M}}(\beta,\lambda'_{S})\times(\beta'_{M},\lambda_{B})$ is possible.
Then, when we substitute $\tilde{W}_{M}$ and $\beta'_{M}$ for $\tilde{W}''_{M}$ and $\beta''_{M}$ in \eqref{C12'}, the relation \eqref{C12'} holds.
Therefore, Corollary \ref{1st} holds.
\end{proof}

\end{document}